\documentclass{article}

 \font\blackboard=msbm10 
 \font\blackboards=msbm7 \font\blackboardss=msbm5
 \newfam\black \textfont\black=\blackboard
 \scriptfont\black=\blackboards \scriptscriptfont\black=\blackboardss
 \def\Bbb#1{{\fam\black\relax#1}}

\def\d{{\rm d}}
\def\mi{{\rm i}}
\def\j{{\rm j}}
\def\arcsinh{\mathop{\rm arcsinh}\nolimits}
\def\det{\mathop{\rm det}\nolimits}
\def\G{\mathop{\Gamma}\nolimits}
\def\Re{\mathop{\rm Re}\nolimits}
\def\e{\mathop{\rm e}\nolimits}
\def\Ai{\mathop{\rm Ai}\nolimits}
\def\Qi{\mathop{\rm Qi}\nolimits}
\def\Res{\mathop{\rm Res}\nolimits}
\def\Tr{\mathop{\rm Tr}\nolimits}

\def\defi{\stackrel{\rm def}{=}}
\newcommand\grsim{\mathrel{\mbox{\lower1ex\mbox{\rlap{$\sim$}\raise1ex\mbox{$>$}}}}}
\newcommand\losim{\mathrel{\mbox{\lower1ex\mbox{\rlap{$\sim$}\raise1ex\mbox{$<$}}}}}

\title{Exercises in exact quantization$^1$}
\author{{\bf Andr\'e Voros} \\
\\
CEA, Service de Physique Th\'eorique de Saclay\\
F-91191 Gif-sur-Yvette CEDEX (France)\\
{ E-mail : {\tt voros@spht.saclay.cea.fr}}\\
\\
and\\
\\
Institut de Math\'ematiques de Jussieu--Chevaleret\\
CNRS UMR 7586\\
Universit\'e Paris 7\\
2 place Jussieu,
F-75251 Paris CEDEX 05 (France)}

\begin{document}
\maketitle

\footnotetext[1]{Corrections and updates (July 2003) indicated by footnotes.}

{\abstract

The formalism of exact 1D quantization is reviewed in detail 
and applied to the spectral study of three concrete 
Schr\"odinger Hamiltonians $[-\d^2/\d q^2 + V(q)]^\pm$ on the half-line 
$\{q>0\}$, with a Dirichlet ($-$) or Neumann ($+$) condition at $q=0$. 
Emphasis is put on the analytical investigation
of the spectral determinants and spectral zeta functions with respect to 
singular perturbation parameters. 
We first discuss the homogeneous potential $V(q)=q^N$ as $N \to +\infty$
vs its (solvable) $N=\infty$ limit (an infinite square well):
useful distinctions are established between regular and singular behaviours
of spectral quantities; various identities among the square-well 
spectral functions are unraveled as limits of finite-$N$ properties.
The second model is the quartic anharmonic oscillator:
its zero-energy spectral determinants 
$\det( -\d^2/\d q^2 + q^4 + v q^2)^\pm$ are explicitly analyzed in detail,
revealing many special values, algebraic identities between Taylor coefficients,
and functional equations of a quartic type
coupled to asymptotic $v \to +\infty$ properties of Airy type.
The third study addresses the potentials $V(q)=q^N+v q^{N/2-1}$ 
of even degree: their zero-energy spectral determinants prove 
computable in closed form, and the generalized eigenvalue problems 
with $v$ as spectral variable admit exact quantization formulae 
which are perfect extensions of the harmonic oscillator case 
(corresponding to $N=2$); these results probably reflect the presence of
supersymmetric$^2$ potentials in the family above.
\footnotetext[2]{In the printed version, we mistakenly used 
``quasi-exactly solvable" for ``supersymmetric" potentials throughout
(the two notions happen to agree for $N=2$ and 6, but not beyond).}
}\bigskip

Exact quantization, or exact WKB analysis, supplies new tools for 
the analytical study of the 1D Schr\"odinger equation, 
now including arbitrary polynomial potentials.
Here we initiate applications of an exact method to miscellaneous
concrete problems and models of analytical interest,
emphasizing exact and asymptotic relations for the spectral determinants
and related spectral zeta functions.

We have chosen three rather different quantum potentials to illustrate
a variety of situations. These have some basic common features 
(besides their required one-dimensional and polynomial nature): 
they are rather simple, with a minimal number of parameters, 
to remain concretely manageable; 
one crucial parameter (discrete or continuous) governs the transition
to a singular limit, creating an interesting dynamical 
and analytical situation;
some uniform principles for tackling those problems can be issued 
at the most general level.

In contrast to earlier studies concerned with individual eigenvalue or
eigenfunction behaviour, we seek
the limiting properties of spectral functions, 
which are symmetric functions of all the eigenvalues at once.
We take a semi-rigorous approach, 
in which we argue a global operational scheme
without claiming absolute completeness in every detail.
\medskip

The paper is organized as follows. 

The introductory Sec.1 gives a detailed survey of the exact tools 
to be used here, while an Appendix collects all the results concerning homogeneous potentials. This is done partly for convenience, 
given the lack of comprehensive reviews
for results that are very scattered in time and place of publication,
and also to clarify some parts of our most recent developments
where localized inconsistencies went undetected.

Then, Sec.2 (Exercise I) considers the family of homogeneous potentials 
$V(q)=|q|^N$ as the degree $N$ tends to $+\infty$. 
This is a most singular problem, 
for which many explicit results are however available beforehand, 
and the limiting problem (an infinite square well) is exactly solved 
by elementary means.
We therefore mainly propose and test some general principles 
of investigation, rather than claim truly new results. 
In particular, we suggest criteria for sorting out regular vs singular
types of limiting behaviour in spectral zeta functions and determinants.
Still, we identify several possibly unnoticed properties and
formulae in the $N=\infty$ limit which arise as regular limits 
of nontrivial finite-$N$ properties.

Sec.3 (Exercise II) deals with the quartic anharmonic oscillator family 
$V(q)=q^4 + v q^2$ which is the most common model
for singular perturbation theory (the free harmonic oscillator emerges
in the $v \to +\infty$ limit). 
We single out a pair of one-parameter
spectral functions for their remarkably numerous and simple explicit
properties: the zero-energy determinants 
$\Qi^\pm (v) \defi \det( -\d^2/\d q^2 + q^4 + v q^2)^\pm$
($+$ corresponds to the even-state sector, $-$ to the odd-state sector).
We present a simple WKB technique allowing to express asymptotic relations
between $v$-dependent determinants such as 
$\det( -\d^2/\d q^2 + q^4 + v q^2)^\pm$ and $\det( -\d^2/\d q^2 + v q^2)^\pm$
when $v \to +\infty$.
Then, practically all the analytical results available for the homogeneous
quartic case ($V(q)=q^4$) have counterparts for the functions $\Qi^\pm$
(and their associated spectral zeta functions), 
while their asymptotic properties remind of the Airy functions (see fig.1). 
Several special values are computable and listed in Table 1 against 
the analogous results for the Airy functions 
and the quartic determinants $\det( -\d^2/\d q^2 + q^4 + \lambda)^\pm$.

The final Sec.4 (Exercise III) gives a complete treatment 
of the similar determinants for a different class of binomial potentials,
$V(q)=q^N+v q^{N/2-1}$ (for $N$ even).
Here the formalism yields fully closed forms 
for the zero-energy spectral determinants:
eq.(\ref{DNMF}) for $\det[ -\d^2/\d q^2 + V(|q|)]^\pm$ 
in terms of Gamma functions (plus non-trivial exponential prefactors)
and, when $N$ is multiple of 4, the still simpler eq.(\ref{DNME}) for the
zero-energy determinant of the potential $V(q)$ itself on the whole real line;
exact quantization formulae follow for the corresponding generalized spectra
(in the $v$ variable: eqs.(\ref{EQNM}) and (\ref{EQNE}) respectively).
A broad generalization of the familiar harmonic-oscillator exact results 
is thus obtained; for $N \equiv 2 \ [{\rm mod} \ 4]$, 
this seems to describe a zero-energy cross-section of the formalism for 
{\sl supersymmetric\/}$^2$ (including some {\sl quasi-exactly solvable\/}) 
models.

Although all three problems rely on the same background formalism,
they can be approached fairly independently from one another. 
Accordingly, there are no global conclusions but each Section carries 
its own concluding remarks.

\section{Introduction}

\subsection{General results on spectral functions \cite{V5,VZ}}
Here, an {\sl admissible\/} spectrum is a purely discrete countable set $\{E_k\}_{k=0,1,2,\cdots}$
with $E_k >0$, $E_k \uparrow +\infty$, such that its partition function 
\begin{equation}
\label{theta}
\theta(t) \defi \sum_{k=0}^\infty \e ^{-tE_k} \qquad \Re(t) > 0 \\
\end{equation}
can be asymptotically expanded in increasing (real) powers $\{t^\rho \}$ for $t\downarrow 0$:
\begin{equation}
\label{tas}
\theta(t) \sim \sum_\rho c_\rho t^\rho  \quad (t\downarrow 0), \quad
\mbox{with } \mu \defi -\min {\{\rho\}} >0 
\mbox{ (the ``growth order").}
\end{equation} 
(In summations etc., Latin indices will systematically mean {\sl integers\/},
and Greek indices mean more general real indices, 
namely real-valued functions over the natural integers,
strictly $\uparrow +\infty$ or $\downarrow -\infty$.)

\subsubsection{Spectral zeta functions \cite{VN,VZ,Ez}} 
Eq.(\ref{tas}) implies that the ``Hurwitz", resp. ``plain"
spectral zeta functions
\begin{equation}
\label{zeta}
Z(s,\lambda) \defi \sum_k (E_k+\lambda)^{-s} \quad
(|\arg(\lambda+E_0)|< \pi-\delta), \quad
\mbox{resp. } Z(s) \defi Z(s,0)
\end{equation}
converge for $\Re(s) > \mu$, that
$Z(s)$ has a meromorphic continuation to all complex $s$, with \cite{P}
\begin{equation}
\label{res}
\mbox{polar set: } \{-\rho\}, \quad \mbox{residue formula: }
\lim_{s \to -\rho} (s+\rho) Z(s) = c_\rho /\G(-\rho),
\end{equation} 
\begin{equation}
\label{TR}
\mbox{``trace identities": for }m \in \Bbb N,\  Z(-m) = (-1)^m m! \, c_m   
\quad (c_m\defi 0 \mbox{ for } m \notin \{\rho\})
\end{equation} 
and similarly for $Z(s,\lambda)$, just by substituting
$\e^{-\lambda t} \theta(t)$ for $\theta(t)$; 
for general $\lambda$, 
the leading trace identity (specially useful for eq.(\ref{Scal}) below) 
is then
\begin{equation}
\label{Tr0}
Z(0,\lambda) = \sum_{0 \le n \le \mu} c_{-n}{(-\lambda)^n \over n!} \equiv
Z(0) + \sum_{1 \le n \le \mu} \Res_{s=n} Z(s) \ {(-\lambda)^n \over n} .
\end{equation} 

\subsubsection{Spectral determinants \cite{V5}}
Since $Z(s,\lambda)$ is regular at $s=0$,
a {\sl spectral determinant\/} can be defined by zeta-regularization, as
\begin{equation}
\label{Det}
D(\lambda) \equiv \det(\hat H + \lambda) \defi 
\exp [- \partial_sZ(s,\lambda)]_{s=0} \ ;
\end{equation}
it is an entire function of order $\mu$ with $\{-E_k\}$ as its set of zeros.
Moreover, amidst all such functions, 
$D(\lambda)$ can be precisely picked out in at least two ways.

$\bullet$ On one side, eq.(\ref{tas}) implies 
a {\sl canonical semiclassical\/} behaviour for $D(\lambda)$, 
\begin{eqnarray}
\label{das}
-\log D(\lambda) &\sim& \sum_\rho c_\rho \Gamma_\rho(\lambda) 
\mbox{ for } \lambda \to +\infty, \qquad 
\Gamma_\rho(\lambda) \defi 
\partial_s \Bigl[{\G(s+\rho) \over \G(s)} \lambda^{-(s+\rho)} \Bigr]_{s=0}
\nonumber \\
\mbox{i.e.,} \quad \Gamma_\rho(\lambda) &=& \Bigl\{ \matrix{ 
\G(\rho) \lambda^{-\rho} & \mbox{if }-\rho \notin  \Bbb N \cr
(-(-\lambda)^m / m!) \bigl(\log\lambda-\sum_{r=1}^m 1/r \bigr)
& \mbox{if }-\rho =m \in \Bbb N } \Bigr.
\end{eqnarray}
are the {\sl only\/} terms allowed; no other type, 
including additive constants ($\propto \lambda^0$), can enter this expansion.

$\bullet$ Independently, 
$D(\lambda)$ is also fully specified by expansions around $\lambda=0$: firstly,
in reference to the {\sl Fredholm determinant\/} $\Delta(\lambda)$
(built as a Weierstrass infinite product):
\begin{eqnarray}
\label{DTay}
D(\lambda) &\equiv& 
\exp \Bigl[ -Z'(0) -\sum_{1 \le n \le \mu}{{\tilde Z}(n) \over n} (-\lambda )^n \Bigr] \,\Delta(\lambda),  \\
\label{DPr}
\mbox{where}\quad \Delta(\lambda)  &\defi&  
\prod_{k=0}^\infty \Bigl( 1+{\lambda \over E_k} \Bigr)
\exp \Bigl[ \sum_{1 \le n \le \mu} {(-\lambda)^n \over n E_k^n} \Bigr] \quad
\mbox{(for all }\lambda), \\
\mbox{and}\quad {\tilde Z}(n) &=& Z(n) \quad \mbox{if $Z(s)$ is regular at $n$,
as when } n>\mu \\
{\tilde Z}(1) &=& \lim_{s \to 1}\Bigl( Z(s) - {c_{-1} \over s-1} \Bigr) \quad
(= Z(1) \mbox{ or its finite part})
\end{eqnarray}
(the general formula for ${\tilde Z}(n)$ on a pole is more contrived 
and required only for $\mu \ge 2$ (\cite{V5}, eq.(4.12)), 
whereas $\mu \le 3/2$ in this work).
Lastly, by way of consequence, the determinants are also characterized 
by these Taylor series (converging for $|\lambda|<E_0$),
\begin{equation}
\label{Dtay}
-\log D(\lambda) = Z'(0) + \sum_{n=1}^\infty {\tilde Z(n) \over n} (-\lambda)^n,
\quad  -\log \Delta(\lambda) = \sum_{n>\mu} {Z(n) \over n} (-\lambda)^n.
\end{equation}
The simplest case is $\mu <1$: then
\begin{equation}
\label{DFr}
\Delta(\lambda) = \prod_{k=0}^\infty \Bigl (1+{\lambda \over E_k} \Bigr) =
\exp \Bigl[ -\sum_{n=1}^\infty {Z(n) \over n} (-\lambda )^n \Bigr] , \quad
D(\lambda) \equiv \e^{-Z'(0)} \Delta(\lambda) .
\end{equation}
\medskip

Two other properties are worth mentioning:

-- if a spectrum $\{ E_k\}$ is dilated to $\{\alpha E_k\}\ (\alpha>0)$,
the spectral functions get rescaled to 
\begin{equation}
\label{Scal}
Z(s,\lambda \vert \alpha) \equiv \alpha^{-s} Z (s,\lambda/\alpha)
\quad \Longrightarrow \quad
D(\lambda \vert \alpha) \equiv \alpha^{Z(0,\lambda/\alpha)} D(\lambda/\alpha)
\end{equation}
(a behaviour hence mainly governed by the leading trace identity,
eq.(\ref{Tr0}));

-- finally, all those results extend to analogous complex spectra \cite{QHS}.

\subsection{1D Schr\"odinger operators with polynomial potentials}

We subsequently specialize to 1D Schr\"odinger equations involving 
a polynomial potential $V(q)$ (adjusted to $V(0)=0$), \cite{S,O}
\begin{equation}
\label{Schr}
(-\d^2/\d q^2\,+[V(q)+\lambda]) \psi = 0, \quad 
V(q) = +q^N+ \mbox{[lower order terms]} .
\end{equation}
We call $\hat H^+$ (resp. $\hat H^-$) the Schr\"odinger operator 
on the half-line $\{q>0\}$
with the Neumann (resp. Dirichlet) boundary condition at $q=0$,
and $\hat H$ the Schr\"odinger operator on the whole line 
with the potential $V(|q|)$, whose spectrum we denote by $\{E_k\}$. 
Then $\{E_k\}_{k\ {\rm even}}$ (resp. $\{E_k\}_{k\ {\rm odd}}$) 
is the spectrum of $\hat H^+$ (resp. $\hat H^-$), 
each one is an admissible spectrum in the previous sense, with
\begin{eqnarray}
\label{Ord}
\mbox{growth order } \mu &=& \textstyle{1 \over 2} + \textstyle{1 \over N} \qquad 
(\mu<1 \mbox{ generically, i.e. for } N>2) , \\
\mbox{exponents }\{\rho\} &=& \{ -\mu +j/N \}_{j=0,1,2,\cdots} ;
\end{eqnarray}
their spectral functions $Z^+,\ D^+$ (resp. $Z^-,\ D^-$) are our basic concern,
and have properties as above (with exceptions in the singular case $N=2$).
However, a few results take a neater or a more regular form 
upon recombined functions instead,
\begin{eqnarray}
\label{SC}
Z \defi Z^+ + Z^-, \quad D \defi D^+ D^- \quad
\mbox{ (spectral functions of $\hat H$)}, \\
\label{SK}
Z^{\rm P} \defi Z^+ - Z^-, \quad D^{\rm P} \defi D^+ / D^- 
\mbox{ (``skew" spectral functions)}.
\end{eqnarray}

\subsubsection{Classical ``spectral" functions \cite{V0}}

The quantum spectral functions of the problem (\ref{Schr}) admit 
natural classical counterparts with parallel properties.
The Weyl--Wigner correspondence, for instance, 
associates the following classical partition function to the quantum one 
of eq.(\ref{theta}),
\begin{equation}
\label{TC}
\theta_{\rm cl}(t) = \int_{\Bbb R ^2}
{\d p\,\d q \over 2 \pi} \e^{-[(p^2+V(|q|)]t} \equiv
{1 \over \sqrt{\pi t}} \int_0^{+\infty} \e^{-V(q)t} \d q ,
\end{equation}
with an expansion coinciding with eq.(\ref{tas}) as long as $\rho \le 0$.
The same Mellin transforms as from the quantum $\theta(t)$ to $Z(s,\lambda)$
then yield
\begin{eqnarray}
\label{ZC} Z_{\rm cl}(s,\lambda) &\defi&
{1 \over \G(s)} \int_0^{+\infty} \theta_{\rm cl}(t) \e^{-\lambda t} t^{s-1} \d t
= {\G(s-1/2) \over \G(s) \sqrt\pi} I_0(s,\lambda) \\
\label{IQ}
\mbox{where} \quad I_q(s,\lambda) &\defi& 
\int_q^{+\infty} (V(q')+\lambda) ^{-s+1/2} \d q' \qquad (\Re(s) >\mu); \\
\label{DC}
\mbox{and} \quad
D_{\rm cl}(\lambda) &\defi& \exp [- \partial_s Z_{\rm cl}(s,\lambda) ] _{s=0},
\end{eqnarray}
with properties induced by eq.(\ref{tas}) similar to the quantum case
(but $D_{\rm cl}(\lambda)$ is not an entire function: branch cuts replace
the discrete zeros of $D(\lambda)$). 

The meromorphic continuation of $I_q(s,\lambda)$ is thus important at $s=0$. 
If (for $\lambda > - \inf V$ initially) we compute the expansion
\begin{equation}
\label{BET}
(V(q) + \lambda)^{-s+1/2} \sim 
\sum_\sigma \beta_\sigma (s) q^{\sigma-Ns} \quad 
{\rm for\ } q \to +\infty \qquad
(\sigma={N \over 2},\ {N \over 2} -1, \cdots) 
\end{equation}
(the $\beta_\sigma$ of course also depend on $\lambda$ and on the potential),
then
\begin{equation}
\label{IA}
I_q(s,\lambda) \sim 
-\sum_\sigma \beta_\sigma(s) {q^{\sigma+1-Ns} \over \sigma+1-Ns} 
\qquad (q \to +\infty) ,
\end{equation}
hence it satisfies 
\begin{equation}
\label{Tres}
\lim_{s \to 0} s I_q(s,\lambda) = \beta_{-1}(0)/N = -Z(0,\lambda)/2;
\end{equation}
$\beta_{-1}(s)$ is actually independent of $\lambda$ except for $N=2$;
the latter value for the residue directly comes from eq.(\ref{ZC}),
making another explicit statement of the trace identity (\ref{Tr0}).

We denote here
\begin{equation}
\label{FPI}
{\mathcal I}_q(\lambda) \defi 
\lim_{s \to 0} \Bigl\{ I_q(s,\lambda) - {\beta_{-1}(0) \over Ns} \Bigr\} \qquad
(\mbox{the {\sl finite part\/} of } I_q(s,\lambda) \mbox{ at } s=0) .
\end{equation}
Now, some of our earlier statements revolving around this quantity \cite{V0,V6}
require corrections in the most general setting $\beta_{-1}(s) \not\equiv 0$.
Finite parts are also to be extracted: first on the expansion (\ref{IA}), giving
\begin{equation}
\label{I0}
{\mathcal I}_q(\lambda) \sim 
-\sum_{\sigma \ne -1} \beta_\sigma(0) {q^{\sigma+1} \over \sigma+1}
-\beta_{-1}(0) \log q +{\partial_s \beta_{-1}(0) \over N} \quad (q \to +\infty),
\end{equation}
then on the definition (\ref{ZC},\ref{DC}), giving
\begin{equation}
\label{SY}
{1 \over 2} \log D_{\rm cl}(\lambda) \defi  -{1 \over 2} \partial_s Z_{\rm cl}(s,\lambda)_{s=0}
= {\mathcal I}_0(\lambda) + 2(1-\log 2) {\beta_{-1}(0) \over N} .
\end{equation}
The finite part ${\mathcal I}_0(\lambda)$ of 
$\int_0^{+\infty} (V(q)+\lambda) ^{-s+1/2} \d q$ at $s=0$ is a candidate
to define a ``symbolic" value for the divergent integral
$\int_0^{+\infty} \Pi_\lambda(q) \d q$, where
\begin{equation}
\label{PI}
\Pi_\lambda(q)  \defi (V(q)+\lambda) ^{1/2},  
\mbox{ the classical (forbidden-region) momentum} .
\end{equation}
However, such an assignment being conventional, 
we much prefer the ``renormalization" given by eq.(\ref{SY}), 
which only adds an explicit constant to the finite part.
So, (more generally) we pose the suggestive notation
\begin{equation}
\label{INTP}
\int_q^{+\infty} \Pi_\lambda(q') \d q' \defi 
{1 \over 2} \log D_{\rm cl}(\lambda) - \int_0^q  \Pi_\lambda(q') \d q' 
\equiv {\mathcal I}_q(\lambda) + 2(1-\log 2) {\beta_{-1}(0) \over N} .
\end{equation}
A big advantage of eq.(\ref{INTP}) will be that 
its $\lambda \to +\infty$ expansion {\sl has the canonical form\/} (\ref{das})
(essentially because $\log D_{\rm cl}$ behaves similarly to its quantum counterpart $\log D$
in this respect, and $\int_0^q  \Pi_\lambda(q') \d q'=O(\sqrt \lambda)+o(1)$ 
is also manifestly canonical).

Thanks to eq.(\ref{I0}), the prescription (\ref{INTP}) 
is also directly characterized by its large-$q$ asymptotic behaviour, as
\begin{eqnarray}
\label{REs}
\int_q^{+\infty} \Pi_\lambda(q') \d q' &=&
-{\mathcal S}_\lambda(q) - \beta_{-1}(0) \log q + {\mathcal C} + o(1) \qquad 
(q \to +\infty) \\
\mbox{where} \quad {\mathcal S}_\lambda(q) &\defi&
\sum_{\{\sigma>-1\}} \beta_\sigma(0) {q^{\sigma+1} \over \sigma+1}, \qquad
{\mathcal C} \defi {1 \over N} 
\Bigl( - 2 \log 2 \, \beta_{-1}(0) 
+ \partial_s \Bigl[ {\beta_{-1}(s) \over 1-2s} \Bigr]_{s=0} \Bigr). \nonumber
\end{eqnarray}

Classical analogs will also arise for $D^+$ and $D^-$ separately 
(eq.(\ref{DPMC}) below).

\subsubsection{Special features of 1D Schr\"odinger determinants \cite{V0,V6}}

The quantum determinants $D^\pm$ for eq.(\ref{Schr}) can be specified 
in two additional ways.
\medskip

$\bullet$ Either spectrum of $\hat H^+$ or $\hat H^-$ obeys a semiclassical
(high-energy Bohr--Sommerfeld) quantization condition of the form
\begin{equation}
\label{BS}
\sum_\rho b_{-\rho}^\pm E_k^{-\rho} \sim k+1/2 \quad 
\mbox{ for } \textstyle{\rm even \atop odd} \ k \to +\infty,
\end{equation}
which implies Euler--Maclaurin continuation formulae for the zeta functions; 
e.g., down to $\Re(s) \ge 0$,
\begin{equation}
\label{EML}
Z^\pm (s) = \lim _{K \to +\infty} 
\Bigl\{ \sum_ {k<K} E_k^{-s} +{1\over 2} E_K^{-s}
-{1\over 2} \sum_{ \{ \rho <0 \} } {\rho \,b_{-\rho}^\pm \over s+\rho}
E_K^{-s-\rho}
\Bigr\} \quad
\mbox{ for } \textstyle{\rm even \atop odd} \ k,K \ .
\end{equation}
The resulting residues are compatible with eq.(\ref{res}) provided
\begin{equation}
\label{COMP}
{b_{-\rho}^\pm \over 2} \equiv {c_\rho^\pm \over \G(1-\rho)} \quad(\rho \ne 0),
\qquad {b_0^\pm \over 2} \pm {1 \over 4} \equiv c_0^\pm \equiv Z^\pm(0) 
\end{equation}
(the latter specifies the $s=0$ trace identities); here, moreover,
\begin{equation}
\label{tr0}
b_{-\rho}^+ \equiv b_{-\rho}^- \mbox{ for all } \rho \le 0 \qquad
\Longrightarrow \qquad Z(0) = b_0^\pm, \quad Z^{\rm P}(0) = 1/2 ;
\end{equation}
then the same trace identity follows for all $Z(s,\lambda)$ and $Z^{\rm P}(s,\lambda)$
by eq.(\ref{Tr0}), 
except for $N=2$ (eq.(\ref{Tr20}) below).

By the same token, $D^\pm(\lambda)$ become directly specifiable  
as {\sl functionals of the spectrum\/}: in the simplest case $\mu<1$,
i.e. $N>2$,
\begin{eqnarray}
\label{EMc}
\log D^\pm(\lambda)  &=& \lim _{K \to +\infty} \Bigl\{ \sum_ {k<K}
\log (E_{k} + \lambda) + {1\over 2} \log (E_{K} + \lambda) \\
&&\qquad -{1 \over 2} \sum_{ \{ \rho <0 \} } 
b_{-\rho}^\pm E_{K}^{-\rho}
\Bigl( \log E_{K}+{1 \over \rho}\Bigr) \Bigr\} \quad
{\rm for\ }k,K\ \textstyle{\rm even \atop  odd}
\nonumber
\end{eqnarray}
(the expansion coefficients $b_{-\rho}^\pm$ being themselves
functions of the spectrum $\{E_k\}$).
\medskip

$\bullet$ $D^\pm(\lambda)$ are also related to an exact solution
$\psi_\lambda(q)$ of eq.(\ref{Schr}) defined by a particular WKB normalization, 
\begin{equation}
\label{WKB}
\psi_\lambda (q) \sim \Pi_\lambda(q)^{-1/2} 
\e^{ \int_q^{+\infty} \Pi_\lambda(q') \d q'} \qquad 
\mbox{for }\Pi_\lambda(q)  \to +\infty ;
\end{equation}
here $\Pi_\lambda(q)$ is the classical momentum function (\ref{PI}),
and the divergent integral $\int_q^{+\infty} \Pi_\lambda(q') \d q'$
is specifically ``renormalized" through eq.(\ref{INTP});
however, eq.(\ref{REs}) also makes $\psi_\lambda(q)$ directly specified
as {\sl the\/} (unique) solution of the Schr\"odinger equation (\ref{Schr})
that decays for $q \to +\infty$ with the precise behaviour
\begin{equation}
\label{Sib}
\psi_\lambda (q) \sim \e^{\mathcal C} q^{-N/4 \, -\beta_{-1}(0)} 
\e^{-{\mathcal S}_\lambda(q)}
(\equiv \e^{\mathcal C} q^{NZ^-(0,\lambda)} \e^{-{\mathcal S}_\lambda(q)}),
\qquad q \to +\infty 
\end{equation}
(the latter form comes from eqs.(\ref{Tres},\ref{tr0})).
Remark: in parallel to eqs.(\ref{WKB},\ref{Sib}),
\begin{eqnarray}
\label{Sib'}
-\psi_\lambda '(q) &\sim& \Pi_\lambda(q)^{+1/2} 
\e^{ \int_q^{+\infty} \Pi_\lambda(q') \d q' } 
\qquad \mbox{for }\Pi_\lambda(q) \to +\infty \\
\bigl( &\sim& \e^{\mathcal C} q^{+N/4 \, -\beta_{-1}(0)} 
\e^{-{\mathcal S}_\lambda(q)}
= \e^{\mathcal C} q^{NZ^+(0,\lambda)} \e^{-{\mathcal S}_\lambda(q)} 
\quad \mbox{for }q \to +\infty \bigr) . \nonumber
\end{eqnarray}

We refer to $\psi_\lambda (q)$ as the ``canonical recessive" solution.
(It is proportional to the ``subdominant" solution of \cite{S}, ch.2,
but only equal to it if $\beta_{-1}(s) \equiv 0$.
The current normalization also differs from \cite{V1}, and from \cite{V0,V6}
--- where it suffers localized inconsistencies. 
The discrepancies, which for $N>2$ involve $\lambda$-independent factors only, 
ultimately cancel out in all results involving only spectral determinants.)
 
The extension of $\psi_\lambda (q)$ to the whole real line, as a solution 
of eq.(\ref{Schr}) with the potential $V(|q|)$,
is shown by integrations to identically satisfy (\cite{V1} Apps.A,D)
\begin{eqnarray}
\label{IDP}
D(\lambda) &\equiv& C  W_\lambda , \qquad  
D^{\rm P}(\lambda) \equiv C^{\rm P} [-\psi'_\lambda(0) / \psi_\lambda(0)], \\
W_\lambda &\defi& \mbox{Wronskian} \{ \psi_\lambda(-q),\psi_\lambda(q) \}
\equiv -2 \psi_\lambda(0) \psi'_\lambda(0) 
\end{eqnarray}
for some constants $C,\ C^{\rm P}$; 
to identify these, we test a characteristic property of 
the spectral determinants: the canonical large-$\lambda$ asymptotics 
(eq.(\ref{das})) of their logarithms down to constants included;
as regards the right-hand sides in eq.(\ref{IDP}), 
the WKB formulae (\ref{WKB},\ref{Sib'}) (good for large $\lambda$)
supply these asymptotic forms,
\begin{equation}
\label{CL}
W_\lambda / 2 \sim \e^{ 2 \int_0^{+\infty} \Pi_\lambda(q) \d q }
\equiv D_{\rm cl} (\lambda), \quad 
 -\psi'_\lambda(0) / \psi_\lambda(0) \sim \Pi_\lambda(0) \quad
(\lambda \to +\infty),
\end{equation}
both of which have canonical ($\lambda \to +\infty$)
logarithms ($D_{\rm cl} (\lambda)$ by analogy with $D(\lambda)$,
and $\Pi_\lambda(0)$ by inspection); hence necessarily,
\begin{equation}
\label{ID0}
D(\lambda) \equiv W_\lambda/2 \equiv -\psi_\lambda(0) \psi'_\lambda(0), \qquad
D^{\rm P}(\lambda) \equiv -\psi'_\lambda(0) / \psi_\lambda(0).
\end{equation}
Incidentally, the second eq.(\ref{CL}) then naturally specifies 
a ``classical skew determinant", $D_{\rm cl}^{\rm P}(\lambda) \defi \Pi_\lambda(0)$,
from which classical analogs of $D^\pm(\lambda)$ also follow as
\begin{equation}
\label{DPMC}
D_{\rm cl}^\pm(\lambda ) \defi 
\Pi_\lambda(0)^{\pm 1/2} \e^{\int_0^{+\infty} \Pi_\lambda(q) \d q} .
\end{equation}

The main conclusion however concerns the quantum spectral determinants $D^\pm$
themselves: upon a straightforward simplification of eq.(\ref{ID0}), 
they get expressed in terms of $\psi_\lambda (q)$ by the fundamental identities
\begin{equation}
\label{ID}
D^-(\lambda ) \equiv \psi_\lambda (0), \qquad 
D^+(\lambda ) \equiv -\psi'_\lambda (0) ,
\end{equation}
(also valid for a rescaled potential, i.e., $V(q)= v q^N+ \cdots$).  

\subsubsection{The main functional relation \cite{V0,V6}}
Jointly with the original problem (\ref{Schr}), 
its set of ``conjugate" equations is defined by means of complex rotations, as
(\cite{S}, ch.2)
\begin{equation}
\label{Cnj}
V^{[\ell]}(q) \defi \e^{-\mi\ell\varphi} V(\e^{-\mi\ell\varphi/2}q), \quad
\lambda^{[\ell]} \defi \e^{-\mi\ell\varphi} \lambda, \quad
\varphi \defi {4\pi \over N+2}
\end{equation}
where $\ell=0,1,\cdots,L-1 \ [{\rm mod}\ L]$ labels the distinct conjugates, 
of total number
\begin{equation}
\label{ELL}
L=N+2 \quad \mbox{in general}, \qquad
L={N \over 2}+1 \quad \mbox{for even polynomial potentials.}
\end{equation}

The main result to be used throughout is the {\sl Wronskian identity\/},
which states a bilinear functional relation between the spectral determinants 
$D^\pm(\lambda )$ and those of the first conjugate equation,
namely $D^{[1]\pm}(\e^{-\mi\varphi} \lambda)$:
\begin{equation}
\label{DW}
\e^{+\mi\varphi/4} D^{[1]+}( \e^{-\mi\varphi} \lambda)
D^{[0]-}( \lambda)
-\e^{-\mi\varphi/4} D^{[0]+}( \lambda) 
D^{[1]-}( \e^{-\mi\varphi} \lambda) \equiv 
2 \mi \e^{\mi\varphi\beta_{-1}(0)/2}.
\end{equation}
It entails an {\sl exact quantization formula\/} for the eigenvalues $E_k$,
\begin{equation}
\label{EQ}
{2 \over \pi} \arg D^{[1]\pm}(-\e^{-\mi\varphi} E)_{E=E_k} 
- {\varphi \over \pi} \beta_{-1}(0) =
 k+{1 \over 2} \pm {N-2 \over 2(N+2)} 
\qquad \mbox{for } k= \textstyle{0,2,4,\ldots \atop 1,3,5,\ldots}
\end{equation}
i.e., this condition determines the spectrum $\{E_k\}$ exactly in terms of
the spectrum $\{E_k^{[1]}\}$ of the first conjugate potential $V^{[1]}$, 
(of which the left-hand-side $D^{[1]\pm}(-\e^{-\mi\varphi} E)$ are functionals).
(Eq.(\ref{EQ}) together with all its conjugates appear to form a determined
system for the resolution of all the spectra $\{E_k^{[\ell]}\}$ at once.)

Here we will often invoke the specific results relating to the 
{\sl homogeneous\/} potentials
\begin{equation}
\label{Sch}
\hat H_N \defi -{\d^2 \over \d q^2}+q^N, \quad q\in[0,+\infty) \qquad 
N\ge 1 \mbox{ integer.}
\end{equation}
The corresponding formulae are collated in greater detail separately
in Appendix, and also in Table 1 below for $N=1,\ 4$.

\subsection{Statement of the problem}

The previous exact analysis of Sec. 1.2 is controlled 
in an essential manner by the degree $N$ of the potential, 
quite sensibly since $q^N$ defines the most singular interaction term. 
Still, there are interesting transitional situations
where this parameter $N$ can diverge or behave discontinuously 
while the quantum problem itself has a well-defined limit
(example: the term $q^N$ is multiplied by a coupling constant $g \to 0$).
Earlier studies of such problems show those limits to be very singular,
and non-uniform over the energy range. 
An unexplored challenge is then 
to control the limiting behaviour of spectral functions,
which are symmetric functions involving all eigenvalues simultaneously. 
The earlier analyses, which essentially work at fixed $N$, 
need further development to handle such non-uniform regimes.

Here we will begin to gather some insight about this issue by examining 
three model problems at varying depths.

\section{Exercise I: infinite square well as $N\to +\infty$ limit}

The infinite square-well potential over the interval $[-1,+1]$
can be realized as the $N\to +\infty$ limit of the homogeneous potential 
$V(q) \equiv |q|^N$. The solutions of the Schr\"odinger equation 
with this potential indeed approach those of the infinite square well, 
but the limiting behaviours are interestingly singular, 
and non-uniform with respect to the quantum number $k$ \cite{FB}.

\subsection{The $N = +\infty$ problem}

The limiting Schr\"odinger operator $\hat H_\infty$ is given by
$V(q) \equiv 0$ in $[-1,+1]$ with Dirichlet boundary conditions at $q=\pm 1$.
At infinite $N$, the spectrum of $|q|^N$ becomes explicit again, 
$\{E_k=(k+1)^2 {\pi^2 / 4}\}$, of growth order $\mu_\infty = {1 \over 2}$. 
As with finite $N$, the operator $\hat H_\infty$ splits into 
$\hat H_\infty^+$ (over the even eigenfunctions, labeled by even $k$)
and $\hat H_\infty^-$ (over the odd eigenfunctions, labeled by odd $k$).

Here the immediately explicit spectral functions are
the {\sl Fredholm\/} determinants $\Delta^\pm$ of eq.(\ref{DFr})
(they reduce to standard Weierstrass products),
\begin{equation}
\label{Frinf}
\Delta_\infty^+(\lambda) = \cos \sqrt{-\lambda}, \quad 
\Delta_\infty^-(\lambda) = {\sin \sqrt{-\lambda} \over \sqrt{-\lambda}}
\quad \Longrightarrow \quad 
\Delta_\infty(\lambda) = {\sin 2\sqrt{-\lambda} \over 2\sqrt{-\lambda}}
\end{equation}
and the plain spectral zeta functions,
\begin{equation}
\label{Zinf}
Z_\infty^+(s) = {2^{2s}-1 \over \pi^{2s}} \zeta(2s), \quad
Z_\infty^-(s) = {1 \over \pi^{2s}} \zeta(2s)
\quad \Longrightarrow \quad
Z_\infty(s) = \Bigl( {2 \over \pi} \Bigr) ^{2s} \zeta(2s).
\end{equation}
The latter formulae imply the explicit computability of $Z_\infty^\pm(s)$ 
at all integers $s \in \Bbb Z$, e.g.,
\begin{equation}
\label{Zi0}
Z_\infty^+(0) = 0, \quad Z_\infty^-(0) = -1/2
\quad \Longrightarrow \quad Z_\infty(0) =  -1/2,
\end{equation}
and ultimately that of the spectral determinants themselves
(even though these will not serve here) through 
$D_\infty^\pm (\lambda) = 
\exp [-(Z_\infty^\pm)'(0)] \Delta_\infty^\pm (\lambda)$, using
\begin{equation}
\label{Di0}
\exp [-(Z_\infty^\pm)'(0)] \equiv D_\infty^\pm (0) =2
\quad \Longrightarrow \quad  D_\infty(0)=4.
\end{equation}

\subsection{The transitional behaviour problem}

Regarding the behaviour of the spectral functions, 
a first task is to seek conditions ensuring the regular behaviour of a quantity
(meaning that it has a finite limit, {\sl and\/} this is the correct value
for the limiting problem). For the other (singular or pathological) quantities
then comes the additional task of describing their precise behaviours.

A fundamental quantity in these problems is the growth order $\mu$.
Here, the limiting (square-well) value $\mu_\infty = {1 \over 2}$ 
agrees with the limit of $\mu_N$ ($={1 \over 2}+{1 \over N}$ by eq.(\ref{Ord})); 
hence $\mu$ behaves regularly (as opposed to the later examples).

As a crude dividing line between (generic) singular and regular behaviours, 
we expect that essentially those quantities which converge both for finite $N$ 
{\sl and\/} in the limiting problem should be regular, in particular:
\begin{equation}
\label{REG}
Z_N^\pm(s,\lambda) \to Z_\infty^\pm (s,\lambda) \quad 
\mbox{ iff } \Re (s) > \mu_\infty .
\end{equation}
Singular behaviour should then set in at $\Re (s) = \mu_\infty$ and,
plausibly, become worse as $\Re (s)$ decreases further.
This purely qualitative argument cannot however predict the precise behaviour  
of any singular quantity.

Fortunately, quantitative statements are made easier here
by a set of explicit results for the finite-$N$ problems (cf. Appendix A.1) 
and their counterparts for $N=+\infty$ (Sec.2.1 above). 
Those data indeed behave consistently with the prediction (\ref{REG}) 
taken with $\mu_\infty = {1 \over 2}$.
As examples of regular ($s>\mu_\infty$) behaviours,
$Z_N^\pm(1) \to Z_\infty^\pm(1) = {1/2 \atop 1/6}$ , 
and likewise for the higher-order {\sl sum rules\/} involving $s=2,\ 3,\cdots$
(Sec.2.5 below).
As opposite examples (involving $s <\mu_\infty$):
$Z_N(0) \equiv 0$ for all $N$ whereas $Z_\infty (0)=-1/2$ 
by eqs.(\ref{Z0},\ref{Zi0});
and worse, eq.(\ref{D0}) implies
\begin{equation}
\label{D0s}
\exp [-( Z_N^\pm)' (s=0)] = D_N^\pm(\lambda=0) \sim (N/\pi)^{1/2} \to \infty,
\end{equation}
even though $D_\infty^\pm(0)$ have finite values, 
perfectly defined by eq.(\ref{Di0}), in the $N=+\infty$ problem~!
(All that shows how carefully such transitional problems must be handled.
--- The same formulae show the {\sl skew\/} spectral functions (\ref{SK}) 
to behave slightly better: e.g.,
$Z_N^{\rm P}(0) \equiv 1/2 = Z_\infty^{\rm P}(0)$,
$D_N^{\rm P}(0) \to 1 = D_\infty^{\rm P}(0)$.)

We then basically expect
\begin{equation}
\label{Dreg}
\Delta_N^\pm (\lambda) \to \Delta_\infty^\pm (\lambda) \quad
\mbox{ but }D_N^\pm(\lambda) \mbox{ \sl diverge\/}, 
\end{equation}
this divergence being confined here to the factor
$\exp [- Z'_N(0)] = D_N(0)$ alone, because $\mu_\infty<1$ 
and the Fredholm determinants can be expressed using $s=1,2,\cdots$ only
as in eq.(\ref{DFr}).
(The Fredholm determinants $\Delta(\lambda)$ should behave regularly 
{\sl in general\/} since they are designed by retaining only {\sl regular\/} values $Z(s)$ (at integers $s$) in their Taylor series (cf. eq.(\ref{Dtay}));
this has to be qualified only if $\mu$ reaches (or jumps across) an integer 
in the limit, as in the example of Sec.3.)

\subsection{The main functional relation}

We now study the $N \to +\infty$ limit of the functional relation (\ref{DW})
specialized to homogeneous potentials $|q|^N$, as stated in eq.(\ref{bfr}).
By the preceding arguments, 
this functional relation should be well-behaved as $N \to +\infty$ 
only once it has been transcribed for Fredholm determinants
(using eqs.(\ref{DFr},\ref{DN0})):
\begin{equation}
\label{Ffr}
\e^{+\mi \varphi/4} \Delta_N^+ (\e^{-\mi \varphi} \lambda) \Delta_N^- (\lambda)
-\e^{-\mi \varphi/4} \Delta_N^+ (\lambda) \Delta_N^- (\e^{-\mi \varphi} \lambda)
\equiv 2 \mi \sin \, \varphi/4 .
\end{equation}
In the latter formula, holding for all $\varphi = 4\pi /(N+2)$,
the expansions of both sides in powers of $\varphi \to 0$ 
should then be identified order by order.
Eq.(\ref{Ffr}) having the form of a ``quantum Wronskian" identity 
for finite $\varphi$ \cite{BLZ,DT},
it is not surprising that the identification to the leading order $O(\varphi)$ 
discloses a ``classical" Wronskian structure:
\begin{equation}
\label{frinf}
\Delta^+ \Bigl(\lambda {\d \over \d\lambda} \Delta^- \Bigr)
- \Bigl(\lambda {\d \over \d\lambda} \Delta^+ \Bigr) \Delta^-
\equiv {1 \over 2} (1- \Delta^+ \Delta^-) .
\end{equation}
However, we know neither how to interpret the right-hand side,
nor how to solve this functional relation directly
(and identification at the next order in $\varphi \propto 1/N$ in eq.(\ref{Ffr})
does not appear to yield any new information either):
this constitutes an interesting open problem, 
since the finite-$N$ equation admits of constructive solutions 
by an exact quantization method using eq.(\ref{EQN});
this method itself however
seems totally singular in the $N \to +\infty$ limit.
At the same time, the Fredholm determinants
of the infinite square well are known, given by eq.(\ref{Frinf});
{\sl they explicitly verify\/} eq.(\ref{frinf}),
and this provides a positive test of regular $N \to +\infty$ behaviour
for the main functional relation in the form (\ref{Ffr}). 

\subsection{The $N=\infty$ coboundary and cocycle identities}

For the homogeneous finite-$N$ problem, a closed functional equation
for the complete determinant $D(\lambda)$ is supplied in Appendix 
as ``the cocycle identity" (\ref{COC}), a sum of $L =O(N)$ terms.
Its naive $N \to +\infty$ limit will be an integral relation for $D(\lambda)$,
further reducible by the residue calculus.
It is however simpler to work out the $N \to +\infty$ limit directly 
upon the (logarithm of the) underlying ``coboundary identity" (\ref{Bfr}) 
equivalent to eq.(\ref{Ffr}): 
this limit is manifestly equivalent to eq.(\ref{frinf}) 
and has the (additive) coboundary form,  
\begin{equation}
\label{COBI}
-2\lambda {\d \over \d\lambda} \,\log \Delta^{\rm P}(\lambda) \equiv 
{1 \over \Delta(\lambda)} - 1 .
\end{equation}
$\Delta^{\rm P}$ being meromorphic and $\Delta$ entire, 
the main solvability condition for eq.(\ref{COBI}) is that the residues at 
the poles of $1/\Delta$ must match the explicit residues of the left-hand side, 
resulting in a curious constraint upon $\Delta$ alone at its zeros,
\begin{equation}
\label{COCI}
-2 \Bigl[\lambda {\d \over \d\lambda} \Delta(\lambda) \Bigr] _{\lambda=-E_k} =
(-1)^k
\qquad \mbox{for all } k \in \Bbb N ,
\end{equation}
which stands as $N=\infty$ counterpart for the cocycle identity (\ref{COC}). 
It is verified by $\Delta_\infty$ but, as with eq.(\ref{frinf}) before, 
we have no idea about other possible solutions.

\subsection{The $N=\infty$ sum rules}

The Taylor series of both sides of eq.(\ref{COBI}) can be expressed 
with the help of eq.(\ref{DFr}), giving
\begin{equation}
\label{SRG}
2 \sum_{n=1}^\infty Z^{\rm P}(n) (- \lambda)^{n} \equiv 
\exp \Bigl [ \sum_{m=1}^\infty {Z(m) \over m} (- \lambda)^m  \Bigr ] -1
\equiv \sum_{r=1}^\infty {1 \over r!}
\Bigl [ \sum_{m=1}^\infty {Z(m) \over m} (- \lambda)^m  \Bigr ] ^r.
\end{equation}
As in the finite-$N$ case (eq.(\ref{SRGN})), this acts as a generating identity:
the identification of each power $\lambda^n$ in eq.(\ref{SRG}) yields a
sum rule of order $n$, which here expresses the combination 
$2 Z^{\rm P}(n)-Z(n)/n$ in terms of the lower $Z(m)$, as
\begin{equation}
2 Z^{\rm P}(1)-Z(1) = 0, \qquad 
2 Z^{\rm P}(2)-Z(2)/2 = Z(1)^2/2, \qquad (\mbox{etc.}).
\end{equation}
The regular behaviour of eq.(\ref{Ffr}) implies that these sum rules too
must be the limits of their finite-$N$ counterparts (\ref{SR}) and also be
verified by  $Z_\infty^\pm(s)$.
(Due to the special form (\ref{Zinf}) of $Z_\infty^\pm(s)$, 
these rules amount to equating each Bernoulli number $B_{2n}$
to a certain polynomial in its predecessors.)
\medskip

In conclusion, this case provides a testing ground for ideas and
methods applicable to transitional regimes, however it has not yet yielded 
any new results about the underlying spectral problems themselves. 
Still, we have identified several novel structures in the $N=\infty$ problem
as imprints of nontrivial finite-$N$ features in the $N \to \infty$ limit.
There remains to effectively handle the finite-$N$ problem
as a regular deformation of this $N=\infty$ case, 
but this would probably require answering the various questions we left open.

\section{Exercise II: anharmonic perturbation theory as $N=2$ limit}

We now study the approach towards the other singular limit of the formalism, $N=2$.
It cannot be realized through homogeneous polynomials, 
but a transition from $N=4$ to $N=2$ precisely underlies 
the well known perturbation theory for the anharmonic potentials 
\cite{BW,Si,HM,ZJ}
\begin{equation}
\label{UV}
U_g(q) = q^2 + g q^4 \quad (g \to 0^+), \quad \mbox{or equivalently }
V_v(q) = q^4 + v q^2 \quad (v \to +\infty);
\end{equation}
we recall the basic unitary equivalence between the two operators
\begin{equation}
\label{Sym}
\hat H \defi - \d^2 / \d q^2 + V_v \qquad \mbox{and} \qquad
\sqrt v (- \d^2 / \d q^2 + U_g) 
\mbox{ with }g \equiv v^{-3/2} .
\end{equation}

We denote the $v$-dependent spectral functions of $\hat H^\pm$ as
\begin{equation}
\label{Spf}
Z^\pm(s, \lambda ; v) , \quad Z^\pm(s ; v) \defi Z^\pm(s,0;v) , \quad D^\pm(\lambda ; v) .
\end{equation}

The poles of these zeta functions lie at $s=3/4,\ 1/4,\ -1/4,\cdots$; 
moreover, by a straightforward computation of eq.(\ref{BET}), 
$\beta_{-1}(s) \equiv 0$ for an even quartic potential;
hence the leading trace identities are $v$-independent, as 
\begin{equation}
\label{Tr0v}
Z(0, \lambda ;v) \equiv 0, \qquad Z^{\rm P}(0, \lambda ;v) \equiv 1/2
\qquad \mbox{for all finite } v .
\end{equation}

\subsection{The transition $v \to \infty$}

This transition is now discontinuous: 
$N$ retains the fixed value 4 
for all finite $v$ while it has the (more singular) value $N=2$ at $g=0$.
All related parameters are then singular, especially now
the order: $\mu \equiv {3 \over 4}$ for all finite $v$, vs $\mu_\infty = 1$.

Each eigenvalue of $\hat H$ satisfies 
\begin{equation}
\label{Vinf}
E_k \sim \sqrt v (2k+1), \qquad v \to +\infty,
\end{equation}
but not uniformly in $k$ \cite{HM}, hence it is another matter to find
the behaviour of the corresponding spectral determinants 
$D^\pm(\lambda;v)$ themselves, as entire functions of $\lambda$ and $v$.

Following eq.(\ref{REG}) again, we now expect $Z^\pm(s,\lambda;v)$ 
to be well-behaved as $v \to +\infty$ iff $s>\mu_\infty = 1$. 
For instance, the $s=0$ trace identity (\ref{Tr0v}) behaves singularly: 
for the limiting potential $U_{g=0}(q) = q^2$, eq.(\ref{tr20}) gives
\begin{equation}
\label{Tr20}
Z(0,\lambda) \equiv -\lambda/2 \qquad (\mbox{but still, } 
Z^{\rm P}(0,\lambda) \equiv 1/2) .
\end{equation}
According to eq.(\ref{Dtay}), $(Z^\pm)'(0,\lambda;v)$ and $D^\pm(\lambda;v)$ 
should be singular (as previously),
but now so should $Z^\pm(1,\lambda;v)$ and the resolvent trace 
$\partial_\lambda \log D^\pm(\lambda;v)$ (also involving $s=1$);
thereafter, higher derivatives $(\partial_\lambda)^n \log D^\pm(\lambda;v)$ 
should behave regularly (as they only involve $s = n,\ n+1,\cdots$). 
Understanding the $v \to +\infty$ behaviour of $D^\pm(\lambda;v)$ 
then just requires the control of two (pairs of) functions {\sl of $v$ alone\/},
$(Z^\pm)'(0;v)$ --- or equivalently $D^\pm(0;v)$ --- and $Z^\pm(1;v)$.

From now on we will exclusively deal with $D^\pm(0;v)$,
a problem which entirely resides in the $\{\lambda = 0\}$ plane.
It is technically simpler because the anomaly (\ref{Tr20}) present in 
(and only in) the limiting problem $U(q)=q^2$ vanishes at $\lambda = 0$.
Moreover, these restricted  spectral functions $D^\pm(0;v)$ 
will display many explicit properties making them intriguingly similar 
to spectral determinants of homogeneous potentials.
The analysis of the complete spectral determinant $D^\pm(\lambda;v)$ 
(including the coefficient 
$Z^\pm(1;v)$ as $\partial_\lambda \log D^\pm(\lambda;v)_{\lambda=0}$)
is also under way but will be more involved.

\subsection{``Extraordinairy" spectral functions}

We therefore restrict subsequent attention 
to the following pair of restricted determinants:
\begin{equation}
\label{QI}
\Qi^\pm(v) \defi D^\pm(\lambda =0;v) = \det (-\d^2 / \d q^2+q^4+vq^2)^\pm .
\end{equation}
These entire functions of $v$ will display numerous explicit properties.
(Determinants of general {\sl binomial\/} potentials,
$\det (-\d^2 / \d q^2+q^N+vq^M),\ M<N$, can be handled likewise, cf. Sec.4.)

Two immediate results are: a pair of special values 
(cf. eq.(\ref{D0}) and Table 1$\,^3$),
\begin{equation}
\label{QI0}
\Qi^+ (0) = D_4^+(0) =  {6^{1/3} 2 \sqrt \pi \over \G(1/6)} ,
\qquad \Qi^- (0) = D_4^-(0) = {\G(1/6) \over 6^{1/3} \sqrt \pi} ;
\end{equation}
and a main functional relation,
drawn from eq.(\ref{DW}) (with $\beta_{-1} \equiv 0$)
and from the conjugacy formula 
$V^{[\ell]}(q) \equiv q^4+(\j^\ell v) q^2$ (cf. eq.(\ref{Cnj})):
\begin{equation}
\label{QIfr}
\e^{+\mi\pi/6} \Qi^+(\j v) \Qi^-(v) - \e^{-\mi\pi/6} \Qi^+(v) \Qi^-(\j v) 
\equiv  2 \mi\ \qquad (\j \defi \e^{2\mi\pi/3}).
\end{equation}
This functional relation is identical to 
that of the homogeneous quartic problem (\ref{bfr4})
if the even and odd arguments are interchanged, 
otherwise it also resembles the Airy relation (\ref{bfr1}). 
Actually, $\Qi^\pm$ will display hybrid properties between those two cases
(hence their name).

\begin{table}
\begin{tabular}  {lllllll}
\hline
& & & & & & \\[-8pt]
Spectra: & \hfill Airy & zeros & \hfill quartic & oscillator &
 \hfill zeros & of $\quad \Qi^\pm$ \\[4pt]
$Z(s)$: & ~~~~~$Z_1^+$  & ~~~~~$Z_1^-$ &
 ~~~~~$Z_4^+$ & ~~~~~$Z_4^-$ &
 ~~~~~${\mathcal Z}^+$ & ~~~~~${\mathcal Z}^-$ \\[2pt]
\hline
& & & & & & \\[-6pt]
$Z'(0)$ & 0.0861126 & $-$0.2299537 & $-$0.1460318 & $-$0.5471153 &
 $-$0.1685422 & $-$0.1780313 \\[2pt]
$\e^{-Z'(0)}$ & ${2 \sqrt\pi \over 3^{1/3} \G(1/3)}$ & 
${2 \sqrt\pi \over 3^{2/3} \G(2/3)}$ & 
${2^{4/3} 3^{1/3} \sqrt \pi \over \G(1/6)}$ & 
${2^{2/3} \sqrt \pi \over 3^{1/3} \G(5/6)}$ & 
${6^{1/3}\G(1/4) \over \G(1/6)}$ & ${\sqrt 2 \G(1/6) \over 6^{1/3}\G(1/4)}$ \\
 &  $=D_1^+(0)$ & $=D_1^-(0)$ & $=D_4^+(0)$ &$=D_4^-(0)$ &
 $={\mathcal D}^+(0)$ & $={\mathcal D}^-(0)$ \\
 &  0.9174909 & 1.2585417 & 1.1572330 & 1.7282604 &
 1.1835782 & 1.1948628 \\[8pt]
$Z(0)$ & 1/4 & $-1/4$ & 1/4 & $-1/4$ &
 1/8 & $-1/8$ \\[4pt]
$Z(1)$ &     0      & $ -\tau $ & $=2\, Z_4^-(1)$ & 
${2^{4/3} \pi^3 \over 3^{17/6} \G(2/3)^5}$ & 
$-{3^{4/3}\G(2/3)^5 \over 2^{10/3}\pi^2}$ & $=2 \,{\mathcal Z}^+ (1)$ \\
 &          & $-$0.7290111    & 1.5266059 & 0.7633029 &
 $-$0.1980209 & $-$0.3960418 \\[4pt]
$Z(2)$ & $1/\tau$ & $\tau^2$ & 
\multicolumn{2}{c}{$\scriptstyle{\rm cf.\ \cite{V1}\ App.C,\ eqs.(C.28,33,34)}$}
 &  &   \\
 & 1.3717212 & 0.5314572 & 0.9147383 & 0.0815825 & 
$\approx 0.3578$ & $\approx 0.2377$ \\[4pt]
$Z(3)$ &    1     & $-\tau^3+{1 \over 2}$ &  &  &  &  \\
 &       & 0.1125618 & 0.8414950 & 0.0190222 & 
$\approx 0.10338$ & $\approx 0.03859$ \\\hline
\end{tabular}
\caption{Analytical and numerical zeta-function values for several spectra.
First two columns: Airy zeros$^3$ (cf. Appendix A.2.2; 
notation as in eq.(\ref{TAU}):
$\tau \defi  -\Ai'(0)/\Ai(0) = 3^{1/3}\G(2/3)/ \G(1/3)$). 
Middle two columns: quartic oscillator ($\hat H_4$) levels (cf. Appendix A.2.1).
Last two columns: zeros of $\Qi^\pm$ (cf. Sec.3.2; we could only 
compute ${\mathcal Z}^\pm(2)$ and ${\mathcal Z}^\pm(3)$ numerically 
from scarce spectral data, at an unreliable accuracy).
Computations used analytical formulae when available, 
but were also cross-checked against direct calculations upon numerical spectra.
Note: 
$\G(1/6)=2^{2/3} \sqrt\pi \G(1/3)/\G(2/3) = 2^{2/3} 3^{1/3} \sqrt\pi /\tau$.
}
\end{table}

\footnotetext[3]{A copying error made three numerical values for the Airy case
(at $s=0$) slightly inaccurate in the printed version.}

\subsubsection{Zeros of $\Qi^\pm$}

The entire functions $\Qi^\pm$ vanish at those values $v=-w_k\ (<0)$ 
for which $\lambda=0$ is an eigenvalue for the potential $q^4+vq^2$
({\sl generalized\/} spectral problem). 
As before, the labeling of $w_k$ in increasing order
makes the parity of $k$ match that of the eigenfunctions. 
(We do not expect any complex zeros.)

The first few zeros of $\Qi^\pm$ evaluate as:
\begin{equation}
\label{Qzer}
\matrix{
k & (\Qi^+) & ~~~~~~~~~ & k & (\Qi^-) \cr
0 & -2.2195971 & & 1 & -3.2511776 \cr
2 & -5.4900693 & & 3 & -6.1598396 \cr
4 & -7.9276920 & & 5 & -8.4854215 \cr
6 & -10.029209 & & 7 & -10.525121 \cr
}
\end{equation}
The zeros $(-w_k)$ also obey an {\sl exact quantization condition\/} 
immediately following from eq.(\ref{QIfr}) by analogy with 
its quartic (\ref{EQ4}) and Airy (\ref{EQ1}) counterparts, 
of which it looks as a crossbreed:
\begin{equation}
\label{QEQ}
{2 \over \pi} \arg \Qi^\pm(-\j w)_{w=w_k} = k + {1 \over 2} \pm {1 \over 6}
\qquad \mbox{for } k= \textstyle{0,2,4,\ldots \atop 1,3,5,\ldots}
\end{equation}

Anticipating on the next paragraph on asymptotic results,
we can at once estimate the zeros $\{w_k\}$ for large $k$
by solving the Schr\"odinger equation with potential $V_v(q)=q^4+vq^2$ 
and with $\lambda=0$ semiclassically, for large $v=-w<0$.
For $q \gg 1$, $\psi(q)$ has standard WKB forms 
in both allowed and forbidden regions (decaying in the latter); 
at the same time, for $q \ll \sqrt w$ (far inside the allowed region), 
$\psi(q)$ must approximately satisfy $(-\d^2 / \d q^2 - w q^2) \psi=0$, 
an equation solvable by Bessel functions (cf. eq.(\ref{K})); specifically here,
\begin{equation}
\label{J}
\psi_\pm(q) \propto q^{1/2} J_{\mp 1/4}(\sqrt w \, q^2/2) \quad 
(\textstyle{\rm even \atop odd} \mbox{ solutions}) .
\end{equation}
Matching the two approximations in the intermediate region 
$\{1 \ll q \ll \sqrt w\}$ then yields the semiclassical quantization formula
\begin{equation}
\label{BST}
{1 \over 2\pi} \oint_{p^2+V(q)=0} p \d q \sim k+{3 \over 4} \mbox{ for $k$ even}
, \qquad k + {1 \over 4} \mbox{ for $k$ odd} .
\end{equation}
(Remark: this Bohr--Sommerfeld quantization rule is appropriate 
for a {\sl general\/} symmetric double-well potential 
with the barrier top precisely kept at the energy 0;
it induces a splitting between even and odd quantized actions exactly half-way
between their (quasi)degeneracy towards the bottom of the wells 
and the equidistant spacing achieved high above the barrier top
\cite{BCP}.)

Now, for the present potential $V_{-w}(q)=q^4-wq^2$,
\begin{equation}
\label{S0}
\oint_{p^2+V_{-w}(q)=0}  p \d q = 
2 \int_{-\sqrt w}^{+\sqrt w} \sqrt {w q^2-q^4}\, \d q 
\equiv {4 \over 3} w^{3/2} ,
\end{equation}
giving $\mu={3 \over 2}$ as the growth order for this spectrum.

\subsubsection{Asymptotic properties}

An essential calculation afforded by the present formalism
is the asymptotic evaluation of the spectral determinants for $v \to \infty$. 
The straightforward behaviour (\ref{Vinf}) of the individual eigenvalues
only suggests that $D^\pm(\lambda;v)$ should closely relate to 
$D_2^\pm(\lambda \vert \sqrt v) \equiv \det (-{\d^2 /\d q^2}+vq^2+\lambda)^\pm$
(the operator $-\d^2 / \d q^2 + v q^2$ being equivalent to $\sqrt v \hat H_2$).
We will therefore need the full expressions
of these harmonic spectral determinants for all $v$,
which follow for instance from eqs.(\ref{Scal},\ref{Gam},\ref{tr20}):
\begin{equation}
\label{scal2}
D_2^\pm(\lambda \vert \sqrt v)= 
{(\sqrt 2 v^{1/8})^{\pm 1-\lambda /\sqrt v} \sqrt{2\pi}
\over \G \Bigl( \displaystyle{2 \mp 1 +\lambda/ \sqrt v \over 4} \Bigr)} .
\end{equation}

We will now connect all the determinants 
through the respective canonical recessive solutions (\ref{WKB}),
which can be fully evaluated (at $\lambda =0$) in the WKB approximation: 
$\psi_{4,\lambda=0}$ for the quartic $V_v(q)= q^4 + vq^2$ on the one hand,
for which $\Pi_0 (q) = (q^4 + vq^2)^{1/2}$ and
\begin{equation}
\label{S4}
\int_q^{+\infty} \Pi_0 (q) \d q = 
-{1 \over 3}(q^2+v)^{3/2} \quad \sim 
-{q^3 \over 3}-{v \over 2} q + O({1 \over q}) \quad \mbox{for } q \to +\infty
\end{equation}
\begin{equation}
\label{WKB4}
\Longrightarrow \qquad  \psi_{4,0} \sim (q^4+vq^2)^{-1/4} \e^{-(q^2+v)^{3/2}/3}
\quad \mbox{for } (q^4 + vq^2) \to +\infty ,
\end{equation}
and $\psi_{2,\lambda=0}$ for the harmonic potential $(vq^2)$ on the other hand,
for which $\Pi_0 (q) = v^{1/2} q$ and
\begin{equation}
\label{WKB2}
\int_q^{+\infty} \Pi_0 (q) \d q = -{\sqrt v \over 2} q^2 \quad \Longrightarrow
\quad \psi_{2,0}\sim (vq^2)^{-1/4}\e^{-\sqrt v q^2/2} 
\quad \mbox{for } (vq^2) \to +\infty.
\end{equation}
(In both cases, the large-$q$ behaviour was checked against eq.(\ref{REs}).)

Now comes the central feature when $v \to \infty$: 
while the quartic recessive solution $\psi_{4,0}(q)$ 
(initially specified when $q \to +\infty$) 
has to match a harmonic recessive solution  
upon penetrating the intermediate region $1 \ll q \ll \sqrt v$, 
the {\sl normalization\/} of $\psi_{4,0}$ 
(canonically set for $q \gg \sqrt v$) 
{\sl need not match\/} that of $\psi_{2,0}$ 
(canonically set for $1 \ll q$ {\sl and no $q^4$ term\/}).
A crucial quantity is actually the {\sl ratio\/} $(\psi_{4,0}/\psi_{2,0})$, 
and it simply emerges 
by reexpanding eq.(\ref{WKB4}) for $\sqrt v \gg q$, as
\begin{equation}
\label{D42}
\psi_{4,0} \sim (vq^2)^{-1/4}\e^{-v^{3/2}/3 - \sqrt v q^2/2} 
\sim \e^{-v^{3/2}/3} \psi_{2,0} 
\quad \mbox{for } 1 \ll q \ll \sqrt v .
\end{equation}
Thereupon, invoking eq.(\ref{ID}) once for the quartic case, 
then for the harmonic case at $\lambda =0$, and finally eq.(\ref{scal2}),
the latest result translates to
\begin{equation}
\label{QD2}
D^\pm(0;v) \sim \e^{-v^{3/2}/3} D_2^\pm(0 \vert \sqrt v) 
 \sim \e^{-v^{3/2}/3} v^{\pm 1/8} D_2^\pm(0) , \qquad \mbox{i.e.,}
\end{equation}
\begin{equation}
\label{qas}
\Qi^+(v) \sim {2 \sqrt \pi \over \G(1/4)} v^{+1/8} \e^{-v^{3/2}/3}, \quad
\Qi^-(v) \sim { \sqrt \pi \over \G(3/4)} v^{-1/8} \e^{-v^{3/2}/3} 
\quad (v\to \infty) .
\end{equation}
So, we obtained asymptotic behaviours for these new functions
which strongly resemble those of the Airy functions
($-\Ai'$ and $\Ai$ respectively),
with an identical growth order $\mu=3/2$. 
Similar reasonings should extend eq.(\ref{qas}) 
to complex $v$ with $ |\arg v| <\pi $, and to
\begin{equation}
\label{Qas}
\Qi^+(-w) \sim {4 \sqrt \pi \over \G(1/4)} w^{+1/8} 
\cos \Bigl[{w^{3/2} \over 3} + {\pi \over 8} \Bigr], \quad
\Qi^-(-w) \sim {2 \sqrt \pi \over \G(3/4)} w^{-1/8}
\cos \Bigl[{w^{3/2} \over 3} - {\pi \over 8} \Bigr]
\end{equation}
for $w \to +\infty$.

The asymptotic formula (\ref{BST}--\ref{S0}) for the zeros $(-w_k)$ 
can now be consistently regained:
either from the asymptotic formula (\ref{Qas}) on the negative real axis,
or from the exact quantization condition (\ref{QEQ}) asymptotically expanded
by means of eq.(\ref{qas}) on the half-line $\{ \arg v = \pi/3 \}$.

Fig. 1 plots the pair of functions $\Qi^\pm$ and their asymptotic forms. 
Given their variegated properties, the idea that $\Qi^\pm$ should be reducible
to simpler known functions seems unlikely.

\subsubsection{Spectral functions of the zeros and applications}
We must also consider spectral functions of the generalized spectrum $\{w_k\}$:
its zeta functions ${\mathcal Z}^\pm(s,v)$, 
and spectral determinants ${\mathcal D}^\pm(v)$, 
which now refer to a {\sl singular\/} operator, i.e.,
\begin{equation}
\label{QD}
{\mathcal Z}^\pm(s,v) \equiv \Tr \,[(q^{-1} \hat H_4 q^{-1} + v)^\pm ]^{-s},
\quad  {\mathcal D}^\pm(v) \equiv \det (q^{-1} \hat H_4 q^{-1} + v)^\pm ,
\end{equation}
whereas $\Qi^\pm(v)= \det (\hat H_4 + v q^2)^\pm$;
hence we cannot readily assert that this spectrum $\{w_k\}$ is admissible 
in the sense of Sec.1, 
nevertheless all the ensuing consequences are numerically verifiable 
and support such an assumption.
(If this singular operator $q^{-1} \hat H_4 q^{-1}$ is self-adjoint,
then the generalized spectrum $\{w_k\}$ is real as we assumed.)

Firstly, the semiclassical quantization conditions (\ref{BST}--\ref{S0}) 
for this spectrum fix the leading trace identities according to eq.(\ref{COMP}):
\begin{equation}
\label{QTR}
b_0 ^\pm = \mp 1/4 \qquad \Longrightarrow \qquad {\mathcal Z}^\pm(0) = \pm 1/8 .
\end{equation}

Then, even though $\Qi^\pm(v)$ are spectral determinants (at a frozen energy),
they do not have to coincide with ${\mathcal D}^\pm(v)$.
Simply, both being entire functions with the same order ${3 \over 2}$ 
and the same zeros, they must be related as 
${\mathcal D}^\pm(v) \equiv C^\pm \e^{c^\pm v} \Qi^\pm(v)$ 
($C^\pm,\, c^\pm$ constants).

Quite generally, a complete identification of such free constants 
can be based on a principle of ``semiclassical compliance" whenever
the spectrum of zeros is admissible: 
the spectral determinant is a priori known up to a factor $\exp P(\lambda)$ 
(with $P$ a polynomial of degree $\le \mu=$ the growth order), 
but only for one such $P$ can the {\sl canonical semiclassical form\/}
(\ref{das}) be also satisfied, and this lifts the ambiguity completely.

Here, the asymptotic formulae (\ref{QD2}) are known for $\log \Qi^\pm(v)$
down to the constant terms (included), and only the latter are actually 
non-canonical in $v \to +\infty$, hence necessarily
\begin{equation}
\label{DQ2}
{\mathcal D}^\pm(v) \equiv \Qi^\pm(v) / D_2^\pm(0) .
\end{equation}

Several results follow from eq.(\ref{DQ2}) (and (\ref{QI0})):

\noindent $\bullet$ Explicit {\sl Stirling constants\/} for the spectrum $\{w_k\}$
(playing the same role as $\sqrt{2\pi}$ for the integers, 
in view of eq.(\ref{EMc}) at $\lambda=0$):
\begin{equation}
\label{DQ0}
{\mathcal D}^\pm(0) \equiv \exp[-{\mathcal Z}^\pm(0)] = D_4^\pm(0) / D_2^\pm(0)
\qquad \mbox{(see Table 1)} .
\end{equation}
$\bullet$ Upon a simple explicit rescaling, 
the expansions (\ref{Dtay}) written for ${\mathcal D}^\pm(v)$ yield
\begin{equation}
\label{QTay}
\Qi^\pm(v) = \Qi^\pm(0) 
\exp \Bigl[- \sum_{n=1}^\infty {{\mathcal Z}^\pm (n) \over n} (-v)^n \Bigr]
\qquad (|v|<w_0) .
\end{equation}
Eqs.(\ref{DQ0},\ref{QTay}) also amount to
\begin{equation}
\label{Qtay}
[-\log \Qi^\pm]^{(n)}(0) = \Bigl\{ \matrix{
(-1)^n (n-1)! \ {\mathcal Z}^\pm (n)
\quad (n \ne 0), \cr
({\mathcal Z}^\pm )'(0) + (Z_2^\pm )'(0) \quad (n=0) .}  \Bigr.
\end{equation}

With these results, we can now draw further consequences 
from the functional relation (\ref{QIfr}) 
and from its coincidence with the homogeneous quartic eq.(\ref{bfr4}) 
up to the exchange of parities. 

\noindent $\bullet$ The complete determinants $D=D^+ D^-$ being unaffected by this interchange,
the quartic cocycle identity (\ref{COC4}) of $D_4$ must remain satisfied 
by its analog, namely the new product function $(\Qi^+ \Qi^-)$.
Hence that functional equation (\ref{COC4}) now shows 
{\sl two completely different entire solutions\/}, 
the former having order $\mu_4 = {3 \over 4}$ and the latter 
$\mu_1 = {3 \over 2}$
(tied by the duality relation $\mu_4^{-1} + \mu_1^{-1} =2$).
(We do not know if still other nontrivial entire solutions may exist.)

$\bullet$ Because of eqs.(\ref{DQ2}--\ref{QTay}),
it is now the spectral zeta functions ${\mathcal Z}^\pm(s)$ which inherit 
sum rules for $s=n \ge 1$, and these have to be the quartic rules (\ref{SR4})
with even/odd arguments swapped, giving
\begin{eqnarray}
\label{SRQ}
{\mathcal Z}^-(1)-2{\mathcal Z}^+(1) &=& 0 \nonumber \\
2{\mathcal Z}^-(2)-{\mathcal Z}^+(2) &=&
3[{\mathcal Z}^-(1)-{\mathcal Z}^+(1)]^2 \\
{\mathcal Z}(3) &=& {\mathcal Z}(1)^3/6-{\mathcal Z}(1){\mathcal Z}(2)/2 
\qquad \mbox{(etc.)} \nonumber 
\end{eqnarray}
(every identity of order $3n$ expresses ${\mathcal Z}(3n)$ 
in terms of the lower ${\mathcal Z}(m)$ in exactly the same form 
as for the quartic zeta value $Z_4(3n)$, cf. eq.(\ref{SR4})).
\medskip

We finally evaluate ${\mathcal Z}^\pm(1)$ in closed form, 
by analogy with the derivation of eq.(\ref{Z1}) for $Z_N^\pm(1)$.
The values ${\mathcal Z}^\pm (1)$ themselves are regularized
quantities, but the first sum rule (\ref{SRQ}) also implies 
$-{\mathcal Z}^+(1)$ = ${\mathcal Z}^{\rm P}(1)$, 
and the latter has the (semi)convergent defining series
$\sum_{k=0}^\infty (-1)^k /w_k$. 
This series can actually be summed
when the $w_k$ are more generally defined (for $N \ge 3$) as the roots of 
$\det ( -\d^2 / \d q^2 + |q|^N - wq^2) = 0$, 
i.e., as the eigenvalues of the {\sl singular\/} operator 
$q^{-1} \hat H_N q^{-1}$ (cf. eq.(\ref{QD})); hence
\begin{equation}
\label{ZPQN}
{\mathcal Z}_N^{\rm P}(1) = \Tr \ \hat P (q^{-1} \hat H_N q^{-1})^{-1}
= \Tr \ \hat P q\hat H_N^{-1} q \quad (\hat P = \mbox{ parity operator});
\end{equation}
then, thanks to the explicit formulae (\ref{K},\ref{Green}) 
giving the kernel of $\hat H_N^{-1}$, and in full parallel with eq.(\ref{ZP1}),
this turns into the explicit integral
\begin{equation}
\label{ZPQ1}
{\mathcal Z}_N^{\rm P}(1) = {4 \nu \over \pi} \sin \nu \pi
\int_0^\infty \bigl[ K_\nu(2 \nu q^{1+N/2}) \bigr] ^2 q^3 \d q, \qquad
\nu = {1 \over N+2}
\end{equation}
i.e., a (convergent) Weber--Schafheitlin integral, which finally gives 
\begin{equation}
\label{ZC1}
{\mathcal Z}_N^{\rm P}(1) = { \sin \nu\pi \over 2 \sqrt\pi} (2\nu)^{2-8\nu}
{\G(3\nu) \G(4\nu) \G(5\nu) \over \G(4\nu+1/2)} .
\end{equation}
For $N=4$, by the first of eqs.(\ref{SRQ}), this also yields the special values
$-{\mathcal Z}^+(1) = -{1 \over 2}{\mathcal Z}^-(1)$ as given in Table 1
(as well as $= -(\log \Qi^+)'(0)= -{1 \over 2}(\log \Qi^-)'(0)$, 
thanks to eq.(\ref{Qtay}) for $n=1$).
\medskip

Table 1 includes some values of these zeta functions ${\mathcal Z}^\pm(s)$. 
We computed them both analytically and by brute force 
(using the data from eqs.(\ref{Qzer},\ref{BST}--\ref{S0}) 
within eqs.(\ref{EML},\ref{EMc})),
and were thus able to numerically check all the above results.

\subsection{A mirror problem without a solution}

The new functions $\Qi^\pm$ algebraically resemble the spectral determinants
of the homogeneous quartic potential,
but share many qualitative and asymptotic properties with the Airy functions, 
especially their order $\mu_1 = {3 \over 2}$. 
This coincidence clearly reflects the $N=1 \leftrightarrow N=4$ duality.
It is then tempting to seek a fourth pair of functions in the empty symmetrical position:
verifying the Airy functional identities (with $+/-$ arguments swapped),
but having the order $\mu_4 = {3 \over 4}$ and the qualitative features 
of the homogeneous quartic determinants. 

However, not only is it difficult to conceive such functions 
around the spectral framework of the linear potential, 
but in fact it is easily shown that a perfect mirror solution cannot exist.
The functions of such a pair should have only negative zeros ($- \varepsilon_k$)
to fully resemble the determinants $D_4^\pm$.
They would also obey the Airy sum rules (\ref{SR1}) 
with the $+/-$ superscripts exchanged, i.e.,
\begin{equation}
Z^-(1) = 0, \qquad Z^+(2) = Z^+(1)^2, \qquad 
\mbox{(etc.),}
\end{equation}
with $ Z^\pm(s) = \sum_k \varepsilon_k^{-s}$
(running over ${\rm even \atop odd}\ k$)
{\sl now also for\/} $s=1$ since $\mu = {3 \over 4}<1$ 
(contrary to the Airy case, where $Z^\pm(1)$ underwent regularization). 
Under this precise circumstance, 
each of the leading putative sum rules above is already broken.
Therefore, we can get no other partner functions to $\Qi^\pm$ in the precise manner described.

(By contrast, this question remains open if we relax the $\varepsilon_k$
to allow negative or complex values, or the order constraint $\mu={3 \over 4}$.)

\subsection{Concluding remarks}

Our study of the functions $\Qi^\pm$ has remained introductory. 
We have explicitly proved neither that their zeros are purely real (negative), 
nor that they form an admissible sequence.
It would also be nice to know their asymptotic expansions 
(\ref{qas},\ref{Qas}) to all orders in $v$ as for the Airy function
(those would also give higher trace identities for ${\mathcal Z}^\pm (-m)$);
and accessorily, to find closed forms for ${\mathcal Z}^\pm (2)$ 
like those for $Z_N^\pm (2)$ (\cite{V1} App.C).
(From a general standpoint, one needs to extend the formalism of Sec.1
to operators $\hat H$ which can be {\sl singular\/} as in eq.(\ref{QD}),
so as to encompass the $\Qi^\pm$ functions;
self-adjointness of these operators should also be ascertained.)

The functions $\Qi^\pm$ have truly revealed hybrid features. 
Their analytical and algebraic properties are undoubtedly quartic, 
while their asymptotic properties are close to the Airy case.
(Unlike the Airy functions, they do not satisfy any obvious differential,
or  linear-difference, equations.)
All in all, their structure is very simple and strongly reminiscent of
the spectral determinants of homogeneous potentials (cf. Appendix), 
but overall not reducible to the latter;
thus, $\Qi^\pm$ provide seemingly new solutions to the functional identities
governing the homogeneous quartic determinants.
We therefore hope that they might also find some roles 
in the correspondences recently unraveled between those functional equations 
and exactly solvable models of statistical mechanics or conformal field theory
\cite{DT}.

\section{Exercise III: Supersymmetric$^2$ binomial potentials}

Keeping the same techniques as previously initiated for the functions $\Qi^\pm$,
we now turn to the zero-energy determinants 
for some other binomial potentials on the half-line $\{q \ge 0\}$: namely,
\begin{equation}
\label{QES}
V(q)=q^N + v q^M, \quad \mbox{with } N \mbox{ even and } 
M \equiv {\textstyle{N\over 2}} -1 \mbox{ throughout},
\end{equation}
because the corresponding Schr\"odinger equation (\ref{Schr}) 
has special properties: it is solvable at $\lambda=0$ 
(in terms of confluent hypergeometric functions \cite{E,AS}), 
and for selected values of $v$ it also provides the simplest examples of 
{\sl supersymmetric\/}$^2$ and of {\sl quasi-exactly solvable\/} systems 
\cite{TUS}.
Then all calculations may strictly follow the previous pattern
(referring to Sec.3 for details), 
yet some of the final results will be quite different.

For the exact formalism of Sec.1, retaining the notations of eq.(\ref{Cnj}),
these potentials enjoy an exclusive symmetry, 
namely, all their conjugate potentials are {\sl real\/}:
\begin{equation}
\label{CNJ}
V^{[\ell]}(q) \equiv q^N + (-1)^\ell v q^M \qquad \mbox{for\ all\ } \ell .
\end{equation}
Moreover, the evaluation of eq.(\ref{BET}) gives a special {\sl non-zero\/} 
residue formula for the first time:
$(V(q)+\lambda)^{-s+1/2} \sim q^{N(-s+1/2)} + v(-s+1/2) q^{-1-Ns}$ 
for $q \to +\infty$ implies
\begin{equation}
\label{RESNM}
\beta_{-1}(s) \equiv v(-s+1/2) \qquad (\mbox{independent of } \lambda,\,N) .
\end{equation}

We henceforth focus upon the restricted determinants $D_N^\pm(\lambda=0;v)$,
where
\begin{equation}
\label{DNM}
D_N^\pm(\lambda ; v) \defi 
\det( -\d^2/\d q^2 + q^N + v q^{{N \over 2} -1} + \lambda)^\pm .
\end{equation}

As with $\Qi^\pm$ before, two immediate results are a pair of special values 
explicitly recoverable from eq.(\ref{D0}),
\begin{equation}
\label{DNM0}
D_N^\pm(0 ; 0) = \det ( -\d^2/\d q^2 + q^N)^\pm = D_N^\pm(0) ,
\end{equation}
and a main functional relation, drawn from eq.(\ref{DW})) 
but now taking a special form,
due to the particular dependence of the degree $M$ upon $N$ 
and to eq.(\ref{RESNM}) (we recall that $\varphi = {4\pi  \over N+2}$):
\begin{equation}
\label{BfrNM}
\e^{+\mi\varphi/4} D_N^+(0;-v) D_N^-(0;v) 
- \e^{-\mi\varphi/4} D_N^+(0;v) D_N^-(0;-v)
\equiv 2 \mi \e^{+\mi\varphi v/4} .
\end{equation}

\subsection{Zeros of $D_N^\pm(0;v)$}

The zeros of $D_N^\pm(0 ; v)$ are again the values $v=-w_k<0$ for which 
$\lambda=0$ is an eigenvalue of $\hat H_N +v |q|^M$ 
(a generalized spectral problem).
But now eq.(\ref{BfrNM}) is very close to 
the {\sl harmonic\/} functional relation (\ref{bfr2}),
especially considering the value $\pi$ of the rotation angle 
acting on the spectral variable (and also the special right-hand-side phase);
then just as eq.(\ref{bfr2}), 
eq.(\ref{BfrNM}) splits into real and imaginary parts and reduces to
\begin{equation}
\label{TRF}
D_N^+(0;v) D_N^-(0;-v) \equiv 
2 \bigl(\sin {\varphi \over 2} \bigr)^{-1} \cos \,{\varphi \over 4} (v-1) ,
\end{equation}
which is exactly a (shifted) {\sl Gamma-function reflection formula\/}.
This spectral problem is then exactly solvable like the harmonic case: 
the zeros of the right-hand side, 
which form one doubly infinite arithmetic progression, must simply be dispatched
according to their signs towards one or the other factor on the left-hand side,
resulting in the {\sl exact\/} eigenvalue formulae
\begin{equation}
w_{2n}={N \over 2}+ (N+2)n, \quad w_{2n+1}={N \over 2}+2 + (N+2)n, \quad i.e.,
\end{equation}
\begin{equation}
\label{EQNM}
{2 \over N+2} w_k = k+{1 \over 2} \pm {N-2 \over 2(N+2)}
\qquad \mbox{for } k= \textstyle{0,2,4,\ldots \atop 1,3,5,\ldots}
\end{equation}

The spectrum $\{w_k\}$ therefore has growth order {\sl unity\/}, 
irrespective of $N$.
For $N=2$, the generalized spectral problem restores the standard
harmonic oscillator problem $\det ( -\d^2/\d q^2 + q^2+v)=0$.
In reverse, the subsequent results will prove perfect generalizations 
to all even degrees $N$ of the classic harmonic oscillator properties.
This clearly provides another view on the partial solvability properties 
of the potentials (\ref{QES}).

For instance, we can show that {\sl semiclassical quantization is exact\/}
for this zero-energy generalized spectrum. 
We proceed just as for $\Qi^\pm$ in Sec.3.2.1; for large $v=-w<0$,
now the comparison equation is $( -\d^2/\d q^2 -w q^M)\psi=0$, solved by
\begin{equation}
\label{JM}
\psi_\pm(q) \propto q^{1/2} J_{\mp 2\nu} 
\Bigl(\sqrt w {q^{{N \over 4}+{1 \over 2}} \over {N \over 4}+{1 \over 2}} \Bigr)
\quad ({\textstyle{\rm even \atop odd}} \mbox{ solutions}) ,
\quad \nu \defi {1 \over N+2},
\end{equation}
hence the asymptotic matching in the intermediate region $1 \ll q \ll w^{2\nu}$
yields the semiclassical quantization formula
\begin{equation}
{1 \over 2\pi} \oint_{p^2+V(q)=0} p \d q 
\sim k+{1 \over 2} \pm {N-2 \over 2(N+2)} \quad 
\mbox{for } k \ \textstyle{\rm even \atop odd}
\end{equation}
while the left-hand side, under the change of variables 
$wx = q^{{N \over 2}+1}$, yields
\begin{equation}
\label{S0NM}
{4 \over 2\pi} \int_0^{w^{2\nu}} (w q^M - q^N)^{1/2} \d q =
{4 w \over (N+2)\pi} \int_0^1 \sqrt{1-x \over x} \d x = {2 \over N+2} w ;
\end{equation}
thus the resulting Bohr--Sommerfeld rule 
manifestly coincides with the exact one (\ref{EQNM}).

The exact quantization formula (\ref{EQNM}) --- like our earlier eq.(\ref{QEQ})
for the zeros of $\Qi^\pm(v) = \det ( -\d^2/\d q^2 + q^4 + v q^2)^\pm$ --- 
displays a hybrid character.
The right-hand side of eq.(\ref{EQNM}) is that 
of the {\sl exact\/} quantization formula (\ref{EQ}) 
for an {\sl arbitrary\/} potential of degree $N$
(whose semiclassical quantization formula is definitely different);
whereas its left-hand side is {\sl linear\/} in the spectral variable
(entirely coming from the term $(-\varphi \beta_{-1}(0)/\pi)$
in eq.(\ref{EQ})), {\sl and\/}
its semiclassical form is {\sl exact\/}, both as in the harmonic case.

\subsection{Asymptotic properties}

We can get the asymptotic $v \to +\infty$ form of $D_N^\pm(0;v)$ just as 
in Sec.3.2.2: we relate these determinants to the canonical recessive solution 
$\Psi_N$ for the potential $V(q)=q^N+vq^M$ at $\lambda=0$, 
then match $\Psi_N$ 
with the solution $\psi_M$ of the comparison potential $vq^M$.
Hence we need the canonical WKB forms (\ref{WKB}) for those two potentials, 
and primarily the symbolic integral $\int_q^\infty V(q')^{1/2} \d q'$.

$\bullet$ Full potential $V$: with a change of variables as above, 
$\int V(q)^{1/2} \d q = 4 \nu v \int \sqrt{1+x^2} \ \d x$,
we have the obvious primitive
\begin{eqnarray}
\label{INT}
\int_0^q V(q')^{1/2} \d q' &=& 
2 \nu \Bigl[ v \arcsinh { q^{{N \over 4} + {1 \over 2}} \over \sqrt v} 
+ q^{{N \over 4} + {1 \over 2}}  \bigl(v+q^{{N \over 2}+1} \bigr)^{1/2} \Bigr] 
\\
&\sim& 2 \nu \Bigl[ v \Bigl(\bigl( {N \over 4} + {1 \over 2} \bigr) \log q 
-{1 \over 2}\log v + \log 2 \Bigr) + q^{{N \over 2}+1} + {v \over 2} \Bigr]
 \quad (q \to +\infty); \nonumber
\end{eqnarray}
we must then set $\int_q^{+\infty} V(q')^{1/2} \d q' \defi 
-\int_0^q V(q')^{1/2} \d q' + C$,
with the target that the large-$q$ behaviour of 
$\Psi_N \sim V(q)^{-1/4} \exp \bigl[\int_q^{+\infty} V(q')^{1/2}\d q' \bigr]$
should obey the canonical eq.(\ref{Sib}), 
now with $\e^{\mathcal C}=2^{-v/N}$ by eqs.(\ref{REs},\ref{RESNM}); i.e.,
\begin{equation}
\label{CAN}
\Psi_N(q) \sim 2^{-v/N} q^{-N/4 \, -v/2} \e^{-2\nu q^{1+N/2}}.
\end{equation}
This adjustment for $C$ according to eqs.(\ref{INT},\ref{CAN})
yields the full WKB specification of the canonical recessive solution, 
when $V(q) \equiv q^N+vq^M$, as
\begin{equation}
\label{WKBNM}
\Psi_N(q) \sim 2^{-v/N} \e^{- \nu v (\log v -1 - 2 \log 2)} V(q)^{-1/4}
\exp \Bigl[-\int_0^q V(q')^{1/2} \d q' \Bigr] \quad (V(q) \to +\infty) .
\end{equation}

$\bullet$ Comparison with the potential $vq^M$: the corresponding 
canonical recessive solution $\psi_M$ is expressible from eq.(\ref{K}) exactly, 
but its WKB form suffices here:
\begin{equation}
\label{WKBM}
\psi_M (q) \sim (vq^M)^{-1/4} \exp \bigl [-\sqrt v q^{{M \over 2}+1} \big 
/ \bigl(\textstyle{M \over 2}+1 \bigr) \bigr].
\end{equation}
Now, the matching with eqs.(\ref{WKBNM},\ref{INT}) reexpanded for $v \gg q$ 
yields the result
\begin{equation}
\Psi_N(q) \sim 2^{-v/N} \e^{-\nu v (\log v -1 - 2 \log 2)} \psi_M (q) \qquad (v \to +\infty) .
\end{equation}
This then translates back to the determinants, by eq.(\ref{ID}), as
\begin{equation}
\det( -{\d^2 \over \d q^2} + q^N + v q^M)^\pm \sim 
2^{-v/N} \e^{-\nu v (\log v -1 -2 \log 2)} 
\det( -{\d^2 \over \d q^2} + v q^M)^\pm .
\end{equation}
But $\det( -\d^2/\d q^2 + v q^M)^\pm \equiv v^{4\nu Z_M^\pm(0)} D_M^\pm (0)$
by eq.(\ref{Scal}), so that finally,
\begin{equation}
D_N^\pm(0;v) \sim \e^{-{v (\log v -1) \over N+2}}
2^{{N-2 \over N(N+2)} v} v^{\pm{1 \over N+2}} D_{{N \over 2} -1}^\pm (0), 
\qquad v \to +\infty .
\end{equation}

\subsection{Spectral functions of the zeros and applications}

As in Sec.3.2.3, we also need the spectral determinants ${\mathcal D}_N^\pm(v)$
built directly for the generalized (even and odd) spectra $\{w_k\}$. 
Since these form the exact (semi-infinite) arithmetic progressions (\ref{EQNM}),
of growth order 1, the answer must now have the exact form
${\mathcal D}_N^\pm(v) \equiv 
C^\pm \e^{c^\pm v}/ \G (\nu(v \mp 1) + 1/2)$.
Compliance with eq.(\ref{das}) then fixes the constants, giving
\begin{equation}
\label{DGAM}
{\mathcal D}_N^\pm(v) \equiv 
{\nu^{\nu(v \mp 1)}\sqrt{2\pi} \over \G (\nu(v \mp 1) + 1/2)}
\end{equation}
(which agrees with the harmonic spectral determinants (\ref{Gam}) for $N=2$).

We can now complete the evaluation of the former determinants $D_N^\pm(0;v)$
themselves, which must have the same zeros as ${\mathcal D}_N^\pm(v)$ and the same order unity, hence necessarily
$D_N^\pm(0;v) \equiv C_\pm \e^{c_\pm v} {\mathcal D}_N^\pm(v)$.
This plus the semiclassical constraint (\ref{das}) again fix the constants, yielding
\begin{equation}
\label{DD}
D_N^\pm(0;v) \equiv 
2^{{N-2 \over N(N+2)} v} D_{{N \over 2} -1}^\pm (0) {\mathcal D}_N^\pm(v) .
\end{equation}
Contrary to eq.(\ref{DQ2}) for the $\Qi^\pm$ case, 
here the latter determinants are known by eq.(\ref{DGAM}),
so we end up with {\sl fully closed forms\/},
\begin{equation}
\label{DNMF}
D_N^+(0;v) \equiv 
-{ 2^{-v/N} (4\nu)^{\nu(v+1) +1/2} \G(-2\nu)
\over \G (\nu(v-1) + 1/2) }, \quad
D_N^-(0;v) \equiv 
{ 2^{-v/N} (4\nu)^{\nu(v-1) +1/2} \G(2\nu)
\over \G (\nu(v+1) + 1/2) }.
\end{equation}
Remark: at $v=0$, the identity (\ref{DD}) specifying the ratios 
$ C_\pm \e^{c_\pm v} = D_N^\pm(0;v) /{\mathcal D}_N^\pm (v)$
simply becomes the duplication formula for $\G(2\nu)$,
a fact which also directly fixes the constants $C_\pm$; 
by contrast, we see no ``cheap" way to obtain the constants $c_\pm$ ---
i.e., $c_+ = c_- = (\log 2) \bigl( 2\nu -{1 \over N} \bigr)
= (\log 2) \bigl( {N-2 \over 2(N+2)} \bigr)$;
in particular, eq.(\ref{TRF}) alone is of no avail in this respect.
\medskip

Alternatively, we can invoke the solvability of this potential at zero energy
to obtain the spectral determinants directly, 
just as we deduced eq.(\ref{D0}) for the homogeneous potentials (cf. Appendix).
Here the canonical recessive solution $\Psi_N$ normalized by eq.(\ref{CAN})
can be expressed indifferently in terms of 
a confluent hypergeometric function $U(a,b,z)$ 
or a Whittaker function $W_{\kappa,\nu}(z)$, as 
(\cite{Ka} eq. 2.273(12))
\begin{eqnarray}
\label{UW}
\Psi_N (q) &\equiv& 
2^{-v/N} (4\nu)^{\nu(v-1)+1/2} \e^{-2\nu q^{1+N/2}} 
U(\nu(v-1)+1/2,\, 1-2\nu,\, 4\nu q^{1+N/2}) \nonumber \\
&\equiv& 2^{-v/N} (4\nu)^{\nu v} q^{-N/4} W_{-\nu v,\nu}(4 \nu q^{1+N/2})
\end{eqnarray}
(the normalization is fixed by reference to the known $q \to +\infty$ forms
of these functions).
In turn, the connection formula 
$U(a,b,z) \equiv \G(1-b)M(a,b,z)/\G(1+a-b)+z^{1-b}\G(b-1)M(1+a-b,2-b,z)/\G(a)$,
together with $M(a,b,0) \equiv 1$, determine the values of 
$\Psi_N (0),\ \Psi'_N (0)$, which finally yield the determinants 
through eq.(\ref{ID}): these then coincide with eq.(\ref{DNMF}) indeed. 
(Conversely, the present method can be viewed as a purely spectral derivation 
of the connection formula for that confluent hypergeometric function.
Again, however, the right reference normalization (\ref{CAN}) of $\Psi_N$ 
has to be expressly fed in.)

(Remark: the knowledge of $\Psi_N (q)$ also yields an expression for 
$\sum (-1)^k/w_k$, in full analogy with eqs.(\ref{ZPQN}--\ref{ZC1}) above, 
but here this only recovers a special case of the known integrals
$\int_0^\infty W_{\kappa,\nu}(z)^2 \d z/z$ (\cite{GR}, eq.7.611(4));
likewise, the sum rules for $\sum (\pm 1)^k/w_k^n$ ($n>1$)
should only reproduce elementary identities among Hurwitz zeta-values here.)

For $N=2$, eq.(\ref{DNMF}) restores 
the ordinary harmonic spectral determinants (\ref{Gam}).
The next even case, $V(q)=q^6+v q^2$, is strongly highlighted in the studies 
on quasi-exactly solvable potentials \cite{TUS}.
However, our present results hold identically irrespective of the parity
of the full potential $q^N+vq^M$, beginning with
the potential $q^4 +v q$ on the half-line $\{q>0\}$, 
so we end with a few remarks about these non-even potentials.

\subsection{The case of non-even potentials}

Our exact results above hold equally well for non-even
potentials $V(q)=q^N+v q^{{N \over 2}-1}$, obtained when $N$ is a multiple of 4,
as for even ones.
Then, as always, the exact zeros $\{w_k\}$ of 
$\det( -\d^2/\d q^2 + q^N + v q^{{N \over 2} -1})^\pm$
as given by eq.(\ref{EQNM}) refer to the potential 
defined on the half-line $\{q>0\}$, with a Neumann/Dirichlet condition at $q=0$
for the $+/-$ parity, 
or equivalently to the singular potential $V(|q|)$ over the whole real line.

For $N$ multiple of 4, an additional exact spectral property is derivable.
We now consider the (non-even) potential $V(q)$ over the whole real line.
The complete spectral determinant ${\bf D}$ for such a potential
of even degree $N$ is not given by $D(\lambda)=D^+(\lambda)D^-(\lambda)$ 
(which corresponds to the even potential $V(|q|)$), 
but in full generality by \cite{V6}
\begin{equation}
{\bf D}(\lambda) \equiv {1 \over 2}[D^+(\lambda)D^{[1+N/2]-}(\lambda) + D^{[1+N/2]+}(\lambda)D^-(\lambda)] .
\end{equation}
In particular, for $V(q)=q^N+v q^{{N \over 2}-1}$ 
with $N \equiv 0 \ [{\rm mod} \ 4]$, by eq.(\ref{CNJ}),
\begin{equation}
\label{DNME}
{\bf D}_N(0 ;v) \equiv  
{1 \over 2}[D_N^+(0 ;v)D_N^-(0 ;-v) + D_N^+(0 ;-v)D_N^-(0 ;v)]
\equiv  {\cos \pi\nu v \over \sin \pi\nu}
\end{equation}
(the explicit end result uses eq.(\ref{DNMF})). 
Thus, the values of $w=-v$
such that the potential $q^N -w q^{{N \over 2}-1}$ 
{\sl on the whole real line\/} has a zero mode are also 
{\sl exactly\/} quantized: they are the zeros of $\cos {\pi \over N+2} v$, i.e.,

 \font\blackboard=msbm10 
 \font\blackboards=msbm7 \font\blackboardss=msbm5
 \newfam\black \textfont\black=\blackboard
 \scriptfont\black=\blackboards \scriptscriptfont\black=\blackboardss
 \def\Bbb#1{{\fam\black\relax#1}}

\begin{equation}
\label{EQNE}
w'_n = (N+2)(n+1/2) \quad (n \in \Bbb Z) \qquad 
\mbox{if } N \equiv 0 \ [{\rm mod} \ 4] ,
\end{equation}
(a spectrum naturally invariant under reflection;
whereas in the even potential case $N \equiv 2 \ [{\rm mod} \ 4]$, 
the spectrum $\{w_k\}$ of eq.(\ref{EQNM}) is recovered).

Curiously, the eigenfunctions (\ref{UW}) corresponding to 
the explicit generalized eigenvalues
(\ref{EQNE}) do not seem to reduce to elementary functions,
while they do so for the generalized eigenvalues (\ref{EQNM});
still, the basis that they form might prove useful
(e.g., the $N=4$ basis for general quartic anharmonic oscillators).
\medskip

In conclusion, a unified analytical formalism has displayed
the quantization of the special zero energy in potentials
of the form $q^N + v q^{{N \over 2}-1}$ (beginning with $N=4$), 
as well as the connection formula for the corresponding 
confluent hypergeometric functions, to be clear generalizations of 
the explicit exact quantization scheme for the harmonic oscillator.
Contrary to the previous example, however,
this analysis treats the problem entirely 
by known functions and does not generate any new ones.

We are grateful to A. Turbiner for suggesting us to apply
exact WKB analysis to quasi-exactly solvable potentials.

\appendix

\section{Appendix: formulae for the potentials $q^N$}

We recapitulate specific results and formulae,
otherwise scattered in many references \cite{S,BOW,BPV,P,VN,VL,V1,VZ,V2,V3,V4},
about the spectral functions 
of the homogeneous Schr\"odinger operators on the half-line,
\begin{equation}
\label{SchN}
\hat H_N \defi -{\d^2 \over \d q^2}+q^N, \quad q\in[0,+\infty) \qquad 
N\ge 1 \mbox{ integer.}
\end{equation}
(Not all $N$-dependences will be systematically stated.) 
As in Sec.1.2, we call $\hat H_N^+$ (resp. $\hat H_N^-$) the operator
with the Neumann (resp. Dirichlet) condition at $q=0$.

Three cases of special interest to us will be, on one side,
the harmonic ($N=2$) and linear ($N=1$) cases,
both of which are describable using known special functions,
as opposed to the non-elementary quartic case ($N=4$).
At the same time, $N=4$ and $N=1$ are both regular cases, 
and dual to each other (they share the same number of conjugates $L=3$),
whereas $N=2$ stands out as a singular (confluent) case (also self-dual)
\cite{V4}.

\subsection{General $N$}

Notations:
\begin{equation}
\label{DN}
D_N^\pm(\lambda) \defi \det(\hat H_N^\pm +\lambda), \qquad
\mbox{and} \quad \mu_N = {N+2 \over 2N},\quad \varphi={4\pi \over N+2}\, ,
\end{equation}
The set of exponents in eq.(\ref{tas}) reduces to 
$\{0\} \cup \{(2n-1) \mu\}_{n=0,1,2,\cdots}$; 
the coefficient of the leading singularity $t^{-\mu}$ is
\begin{equation}
c_{-\mu}^\pm = (2 \sqrt \pi)^{-1} \G(1+1/N) .
\end{equation}
In accordance with eq.(\ref{TR}), 
the {\sl trace identities\/} then have the pattern \cite{P,VN}
\begin{equation}
\label{Z0}
Z_N(0)=0, \qquad  Z_N^{\rm P}(0)=1/2
\end{equation}
\begin{equation}
\label{Zm}
\Bigl \{\matrix{
Z_N(-m) = 0 \mbox{ unless } m= ({1 \over 2} + r) (1+{N \over 2}) \cr
Z_N^{\rm P}(-m) = 0 \mbox{ unless } m= r (1+{N \over 2})}
\mbox{ for $m$ and $r \in \Bbb N$} \Bigr.
\end{equation}

$\bullet$ As all the conjugate potentials $V^{[\ell]}(q)$ (eq.(\ref{Cnj}))
coincide here,
the {\sl main functional relation\/} (\ref{DW}) for the spectral determinants
boils down to
\cite{VZ}
\begin{equation}
\label{bfr}
\e^{+\mi \varphi/4} D_N^+ (\e^{-\mi \varphi} \lambda) D_N^-(\lambda)
-\e^{-\mi \varphi/4} D_N^+ (\lambda) D_N^- (\e^{-\mi \varphi} \lambda) 
\equiv 2 \mi \qquad (N \ne 2)
\end{equation}
or equivalently, to a multiplicative ``coboundary identity"
linking the full and skew determinants, \cite{VL,V1,V3}
\begin{equation}
\label{Bfr}
{D_N^{\rm P}(\lambda) / D_N^{\rm P}(\e^{-\mi\varphi} \lambda)} 
\equiv \e^{\mi( -2 \Phi_N(\lambda) + \varphi /2)}, \qquad 
\Phi_N(\lambda) \defi {\rm arcsin} \, 
[ D_N(\e^{-\mi\varphi}\lambda)D_N(\lambda) ]^{-1/2} , 
\end{equation}
the branch of the arcsin being fixed at $\lambda =0$ 
with the help of eq.(\ref{bfr}):
\begin{equation}
\label{DN0}
D_N(0) = (\sin \,\varphi/4)^{-1} \qquad \Longrightarrow \qquad 
\Phi_N(0) \defi \varphi/4.
\end{equation}

$\bullet$ The {\sl exact quantization condition\/}, drawn from eq.(\ref{bfr}), 
is \cite{V2,V3,V4}
\begin{equation}
\label{EQN}
{2 \over \pi} \arg D_N^{\pm}(-\e^{-\mi\varphi} E)_{E=E_k} =
 k+{1 \over 2} \pm {N-2 \over 2(N+2)}
\qquad \mbox{for } k= \textstyle{0,2,4,\ldots \atop 1,3,5,\ldots}
\quad (N \ne 2).
\end{equation}

$\bullet$ By multiplying together all the conjugates of eq.(\ref{Bfr}), 
a corresponding multiplicative {\sl ``cocycle identity"\/}
(of length $L$ given by eq.(\ref{ELL})) results,
providing a consistency condition upon $D$ alone in the implicit form \cite{VL}
\begin{equation}
\label{COC}
\sum_{\ell=0}^{L-1} \Phi_N (\e^{-\mi\ell\varphi}\lambda) \equiv L \varphi/4 ,
\end{equation}
i.e. an autonomous functional equation,
circularly symmetric of order $L$ (and convertible to a polynomial form)
for the complete determinant $D_N$.

$\bullet$ Some {\sl special values\/} of spectral functions also become explicit
(\cite{VN}; \cite{V1} App.C) thanks to the {\sl solvability\/} 
of the differential equation $(\hat H_N +\lambda)\psi_\lambda(q)=0$ 
at $\lambda=0$, 
specifically here by Bessel functions (\cite{E}, vol.II ch.7.2.8):
the {\sl canonical\/} recessive solution (\ref{WKB}), 
of asymptotic behaviour $\psi_0(q) \sim q^{-N/4}\exp[-q^{1+N/2}/(1+N/2)]$
prescribed by eq.(\ref{Sib}), is 
\begin{equation}
\label{K}
\psi_0(q) \equiv 2 \sqrt{\nu/\pi} \, q^{1/2} K_\nu (2 \nu q^{1+N/2}) \qquad
(\nu \defi {1 \over N+2})
\end{equation}
(normalized by reference to the known $z \to +\infty$ behaviour of $K_\nu(z)$).
In turn, the values $\psi_0(0),\ \psi'_0(0)$ follow from 
$K_\nu(z) =\pi (2 \sin \,\nu\pi)^{-1} [I_{-\nu}(z)-I_\nu(z)]$ 
plus $I_{\pm\nu}(z) \sim (z/2)^{\pm\nu}/\G(\pm\nu+1) \mbox{ for }z \to 0$, 
and finally eq.(\ref{ID}) yields
\begin{equation}
\label{D0}
D_N^+(0) = {\G(1-\nu) \over \nu^{N \nu /2} \sqrt \pi}, \quad
D_N^-(0) = {\G(\nu) \nu^{N \nu /2} \over \sqrt \pi} 
\quad \Longrightarrow \quad D_N(0) = {1 \over \sin \, \nu\pi}  
\end{equation}
(the last result agreeing with eq.(\ref{DN0})).

Eq.(\ref{K}) also generates formulae for $Z_N^\pm(n),\ n=1,2,\cdots$,
by fully specifying a general expression of the 1D Green's function,
\begin{eqnarray}
\label{Green}
\langle q \vert (\hat H + \lambda)^{-1} \vert q'\rangle &=&
W_\lambda^{-1} \,\psi_\lambda(-\min\{q,q'\}) \,\psi_\lambda(\max\{q,q'\})  \\
(W_\lambda &\defi& \mbox{Wronskian} \{ \psi_\lambda(-q),\psi_\lambda(q) \}) 
\end{eqnarray}
when $\hat H = \hat H_N$ and $\lambda=0$ 
(with $W_\lambda \equiv 2 D (\lambda)$ by eq.(\ref{ID0}),
implying $W_0  = 2 (\sin\, \varphi/4)^{-1}$ by eq.(\ref{D0})).
Integral formulae giving $Z_N(n)$ as $\Tr \, (\hat H_N ^{-1})^n$, 
and $Z_N^{\rm P}(n)$ as $\Tr \, \hat P (\hat H_N ^{-1})^n$ ($\hat P \defi$ the
parity operator), thus become explicit. For $n=1$, the latter is the simpler:
\begin{equation}
\label{ZP1}
Z_N^{\rm P}(1) = {4 \nu \over \pi} \sin \nu \pi
\int_0^\infty \bigl[ K_\nu(2 \nu q^{1+N/2}) \bigr] ^2 q\ \d q, \qquad
\nu = {1 \over N+2};
\end{equation}
finally this evaluates by a Weber--Schafheitlin formula 
(\cite{E}, vol.II ch.7.14, eq.(36), and likewise for $Z_N(1)$ \cite{VN}):
\begin{equation}
\label{Z1}
Z_N^{\rm P}(1) = { \sin \nu\pi \over 2 \sqrt\pi} (2\nu)^{2-4\nu}
{\G(\nu) \G(2\nu) \G(3\nu) \over \G(2\nu+1/2)} , \qquad
Z_N(1) = {\tan 2\nu\pi \over \tan \nu\pi} Z_N^{\rm P}(1).
\end{equation}
(Remarks: -- the results (\ref{D0}), (\ref{Z1}) also work for $N=1,\ 2$; 
-- this approach can still handle $Z_N^\pm(2)$, but with final results 
reducible only to $_4F_3$ generalized hypergeometric series: \cite{V1}, App. C.)

$\bullet$ For general integer $n$, by contrast, 
the farthest explicit algebraic result we can reach is a {\sl single\/} identity
(a {\sl sum rule\/}) at the level of each doublet $(Z_N^+(n),Z_N^-(n))$,
and this comes simply by expanding the functional identity (\ref{bfr}) in all powers $\lambda^n$, as \cite{V1}
\begin{equation}
\label{SRGN}
\sin \Bigl[ {\varphi \over 4}+ \sum^{ \infty}_{ n=1} 
\sin {n\varphi \over 2}\ {Z_N^{\rm P}(n) \over n}(-\lambda)^ n \Bigr]
\equiv \exp \Bigl[ Z'_N(0)+ \sum^{ \infty}_{ m=1} 
\cos {m\varphi \over 2}\ {Z_N(m) \over m}(-\lambda)^m \Bigr] ,
\end{equation}
then equating both sides of this generating identity at each order $n$:
at $n=0$ we already obtained $D_N(0) = 1/\sin (\varphi/4)$ 
(eq.(\ref{DN0}) or (\ref{D0}));
then, at higher orders, we get sum rules in the form:
\begin{eqnarray}
\label{SR}
\cot {\varphi \over 4} \sin {\varphi \over 2} \, Z_N^{\rm P}(1)
-\cos {\varphi \over 2} \, Z_N(1) &=& 0 \qquad\qquad\qquad (N \ne 2) \nonumber\\
\cot {\varphi \over 4} \sin {2\varphi \over 2} \, Z_N^{\rm P}(2)
-\cos {2\varphi \over 2} \, Z_N(2) &=& 
[2\cos{\varphi \over 4} \, Z_N^{\rm P}(1)]^2 \\
\cot {\varphi \over 4} \sin {3\varphi \over 2} \, Z_N^{\rm P}(3)
-\cos {3\varphi \over 2} \, Z_N(3) &=&
4 \cos{\varphi \over 2} [3\cos{\varphi \over 2} \, Z_N^{\rm P}(1) Z_N^{\rm P}(2)
- 2\cos^2 {\varphi \over 4} \, Z_N^{\rm P}(1)^3 ]  \nonumber \\
&\vdots& \nonumber \\
\cot {\varphi \over 4} \sin {n\varphi \over 2} \, Z_N^{\rm P}(n)
-\cos {n\varphi \over 2} \, Z_N(n) &=& 
\mbox{a polynomial of }\{Z_N^\pm(m)\}_{1 \le m <n} \nonumber
\end{eqnarray}
(the first line also follows from eq.(\ref{Z1})).

Other results are {\sl resurgence properties\/} of the spectral determinants 
(namely, exact analytical constraints on their $(1/\hbar)$-Borel transforms)
\cite{V1,VZ,DPA}.

\subsection{Special cases}

Out of the preceding results valid for general $N$, 
we only restate those which are specifically needed in the main text, 
or which take a particular form; special and numerical values are in Table 1. 
We will be using $\j \defi \e^{2\mi\pi/3}$.

\subsubsection{$N=4$} 
This case (quartic oscillator) has order $\mu={3 \over 4}$.

Main functional relation (\ref{bfr}):
\begin{equation}
\label{bfr4}
 \e^{+\mi\pi/6} D_4^+(\j^{-1}\lambda)D_4^-(\lambda)
-\e^{-\mi\pi/6} D_4^+(\lambda)D_4^-(\j^{-1}\lambda) \equiv 2\mi 
\qquad (\varphi=2\pi/3).
\end{equation}

Exact quantization condition (\ref{EQN}):
\begin{equation}
\label{EQ4}
{2 \over \pi} \arg D_4^\pm(-\j^2 E)_{E=E_k} = k + {1 \over 2} \pm {1 \over 6}
\qquad \mbox{for } k= \textstyle{0,2,4,\ldots \atop 1,3,5,\ldots}
\end{equation}

Cocycle functional equation (\ref{COC}): its algebraic form is
\begin{equation}
\label{COC4}
D_4(\lambda) D_4(\j\lambda) D_4(\j^2\lambda) \equiv
D_4(\lambda) + D_4(\j\lambda) + D_4(\j^2\lambda) + 2.
\end{equation}

Special values for $D_4^\pm(0),\ Z_4^\pm(1)$: see Table 1 (note that $D_4(0)=2$).

The sum rules (\ref{SR}) can also be written for $N=4$ as
\begin{eqnarray}
\label{SR4}
Z_4^+(1)-2Z_4^-(1) &=& 0  \nonumber \\
2Z_4^+(2)-Z_4^-(2) &=& 3[Z_4^+(1)-Z_4^-(1)]^2 \\
Z_4(3) &=& Z_4(1)^3/6-Z_4(1)Z_4(2)/2 \qquad \mbox{(etc.)}. \nonumber
\end{eqnarray}

(In addition, $Z_4^\pm(s)$ have been asymptotically evaluated 
for $s \to -\infty$, 
by exploiting the resurgence equations for the spectral determinants.
\cite{VN,VZ})

\subsubsection{$N=1$} 
The eigenvalues $E_k$ are the (unsigned) zeros of the Airy functions, 
here meant as $\Ai$ (for odd parity) and $\Ai'$ (for even parity) 
(\cite{AS}, ch.10.4).
A few results nevertheless seem new \cite{V4}. The order is $\mu={3 \over 2}$.

The canonical recessive solution simply relates to the Airy function:
\begin{equation}
\psi_\lambda(q) \equiv 2 \sqrt \pi \Ai(q+\lambda).
\end{equation}

The spectral determinants are then also Airy functions by eq.(\ref{ID}),
\begin{equation}
\label{Ai}
D_1^+(\lambda)=-2 \sqrt \pi \Ai'(\lambda), \quad 
\quad D_1^-(\lambda)=2 \sqrt \pi \Ai(\lambda) \quad \Longrightarrow \quad 
D_1(\lambda)= -2\pi (\Ai^2)'(\lambda) ,
\end{equation}
and their functional relation (\ref{bfr}) boils down to 
the classic Wronskian identity for $\Ai(\cdot)$ and $\Ai(\j^2 \cdot)$:
\begin{equation}
\label{bfr1}
\e^{+\mi\pi/3} D_1^+(\j\lambda)D_1^-(\lambda)
-\e^{-\mi\pi/3} D_1^+(\lambda)D_1^-(\j\lambda) \equiv 2\mi 
\qquad (\varphi=4\pi/3) .
\end{equation}

Exact quantization condition (\ref{EQN}):
\begin{equation}
\label{EQ1}
{2 \over \pi} \arg D_1^\pm(-\j E)_{E=E_k} = k + {1 \over 2} \mp {1 \over 6}
\qquad \mbox{for } k= \textstyle{0,2,4,\ldots \atop 1,3,5,\ldots}
\end{equation}

Cocycle identity: the algebraic form of (\ref{COC}) for $N=1$ is
\begin{equation}
\label{COC1}
D_1(\lambda)^2 + D_1(\j\lambda)^2 + D_1(\j^2\lambda)^2
-2 [ D_1(\j\lambda)D_1(\j^2\lambda) + D_1(\j^2\lambda)D_1(\lambda)
+ D_1(\lambda)D_1(\j\lambda) ] +4 \equiv 0 .
\end{equation}

Eqs.(\ref{D0}) for $D_1^\pm(0)$ just express the known values 
of $\Ai(0),\ \Ai'(0)$ (see Table 1).

Sum rules (\ref{SR}):
\begin{equation}
\label{SR1}
Z_1^+(1) = 0, \quad Z_1^-(2) = Z_1^-(1)^2, \quad
Z_1(3) = 5 Z_1(1)^3/2 - 3 Z_1(1)Z_1(2)/2, \quad \mbox{(etc.)}.
\end{equation}
Here, due to a supplementary set of relations, 
all values $Z_1^\pm(n)$ become recursively expressible as rational functions of 
\begin{equation}
\label{TAU}
\tau \defi D_1^{\rm P}(0) \equiv -\Ai'(0)/\Ai(0) = 3^{1/3}\G(2/3)/ \G(1/3)
\approx 0.729011133 \, ;
\end{equation}
they are all irrational except $Z_1^+(1) = 0$ (a regularized value, however)
and $Z_1^+(3) = 1$ (Table 1 lists all $Z_1^\pm(n)$ analytically up to $n=3$).

\subsubsection{$N=2$} 
{\sl The\/} finite-$N$ case where the spectrum is known ($E_k=2k+1$), 
of order $\mu=1$ (an integer, which makes it a singular case).

The canonical recessive solution relates to a parabolic cylinder function
(\cite{AS}, ch.19; \cite{E}, vol.II ch.8)
\begin{equation}
\label{PC}
\psi_\lambda(q) \equiv 2^{(1-\lambda)/4} U(\lambda/2,\,\sqrt 2\,q).
\end{equation}

The determinants are given by Gamma functions,
\begin{equation}
\label{Gam}
D_2^+(\lambda) = {2^{-\lambda/2}2\sqrt\pi \over 
\G \bigl( {1+\lambda \over 4} \bigr)}, \quad
D_2^-(\lambda) = {2^{-\lambda/2}\sqrt\pi \over 
\G \bigl( {3+\lambda \over 4} \bigr)}
\quad \Longrightarrow \quad 
D_2(\lambda) = {2^{-\lambda/2} \sqrt{2\pi} \over 
\G \bigl( {1+\lambda \over 2} \bigr)};
\end{equation}
their functional relation is now drawn directly from eq.(\ref{DW}) 
instead of (\ref{bfr}):
\begin{equation}
\label{bfr2}
\e^{+\mi\pi/4} D_2^+(-\lambda)D_2^-(\lambda)
-\e^{-\mi\pi/4} D_2^+(\lambda)D_2^-(-\lambda) \equiv 2\mi \e^{\mi\pi\lambda/4}
\qquad (\varphi=\pi),
\end{equation}
and its right-hand side carries a special extra factor
because for homogeneous potentials,
\begin{equation}
\label{Res}
\beta_{-1}(s) \equiv 0, \quad  \mbox{except:} \quad
\beta_{-1}(s) \equiv \lambda (-s+1/2) \ \mbox{ for } N=2.
\end{equation}
For this special value $\varphi=\pi$, the functional relation further splits
into its real and imaginary parts at real $\lambda$, and thereby reduces to 
\begin{equation}
\label{Bfr2}
D_2^+(\lambda)D_2^-(-\lambda) \equiv 2 \cos \ {\pi \over 4}(\lambda -1);
\end{equation}
hence it just amounts to the reflection formula for $\G(z)$
(like the cocycle identity, which we do not write).

Here, the exact quantization condition $E_k=2k+1$ trivially comes 
by dispatching the obvious zeros of the right-hand side of eq.(\ref{Bfr2}): 
the positive ones to $D_2^-(-\lambda)$, the negative ones to $D_2^+(\lambda)$.

The anomalous value (\ref{Res}) accounts for special $N=2$ trace identities:
\begin{equation}
\label{tr20}
Z_2^\pm(0,\lambda) \equiv (-\lambda \pm 1)/4 ;
\end{equation}
the anomaly vanishes at (and only at) $\lambda=0$, 
and indeed the general formulae (\ref{ID},\ref{D0}) for $D_N^\pm(\lambda =0)$
correctly agree with eq.(\ref{Gam}) to give
\begin{equation}
D_2^+(0) = { 2 \sqrt \pi \over \G({1 \over 4})} \approx 0.977741067, \quad
D_2^-(0) = {\sqrt \pi \over \G({3 \over 4})} \approx 1.446409085 \quad \Longrightarrow
\quad D_2(0)=\sqrt 2 .
\end{equation}

The spectral zeta functions are
\begin{equation}
\label{Z2}
Z_2(s) \equiv (1-2^{-s}) \zeta(s), \qquad 
Z_2^{\rm P}(s) \equiv \beta(s) \ 
\bigl(\defi \sum_{k=0}^\infty (-1)^k/(2k+1)^s \bigr) ;
\end{equation}
their sum rules only reproduce the known special values 
$\zeta(2n)$ and $\beta(2n+1)$ (\cite{AS}, ch.23)
\begin{equation}
\label{SR2}
Z_2^{\rm P}(1) = \pi/4, \qquad Z_2(2) = \pi^2/8, \qquad 
Z_2^{\rm P}(3) = \pi^3/32, \qquad \mbox{(etc.)} .
\end{equation}

\bigskip

\centerline{\bf Figure caption}
\bigskip

The functions $\Qi^-$ and $(-\Qi^+)$ (of eq.(\ref{QI})), which specially
resemble the Airy functions $\Ai$ and  $\Ai'$ respectively, 
plus logarithmic plots to magnify their large-$v$ asymptotics on the decaying
side ($v \gg 0$). 
Numerical values: {\tt +} for $\Qi^+$, and {\tt o} for $\Qi^-$
(computed using eq.(\ref{EMc}),
with an input of $K \approx 10^3$ numerical eigenvalues 
of the operator $(- \d^2 / \d q^2 + q^4 + v q^2)$
for every value of the coupling constant $v$).
The dashed lines plot the large-$v$ asymptotic formulae (\ref{qas},\ref{Qas}).

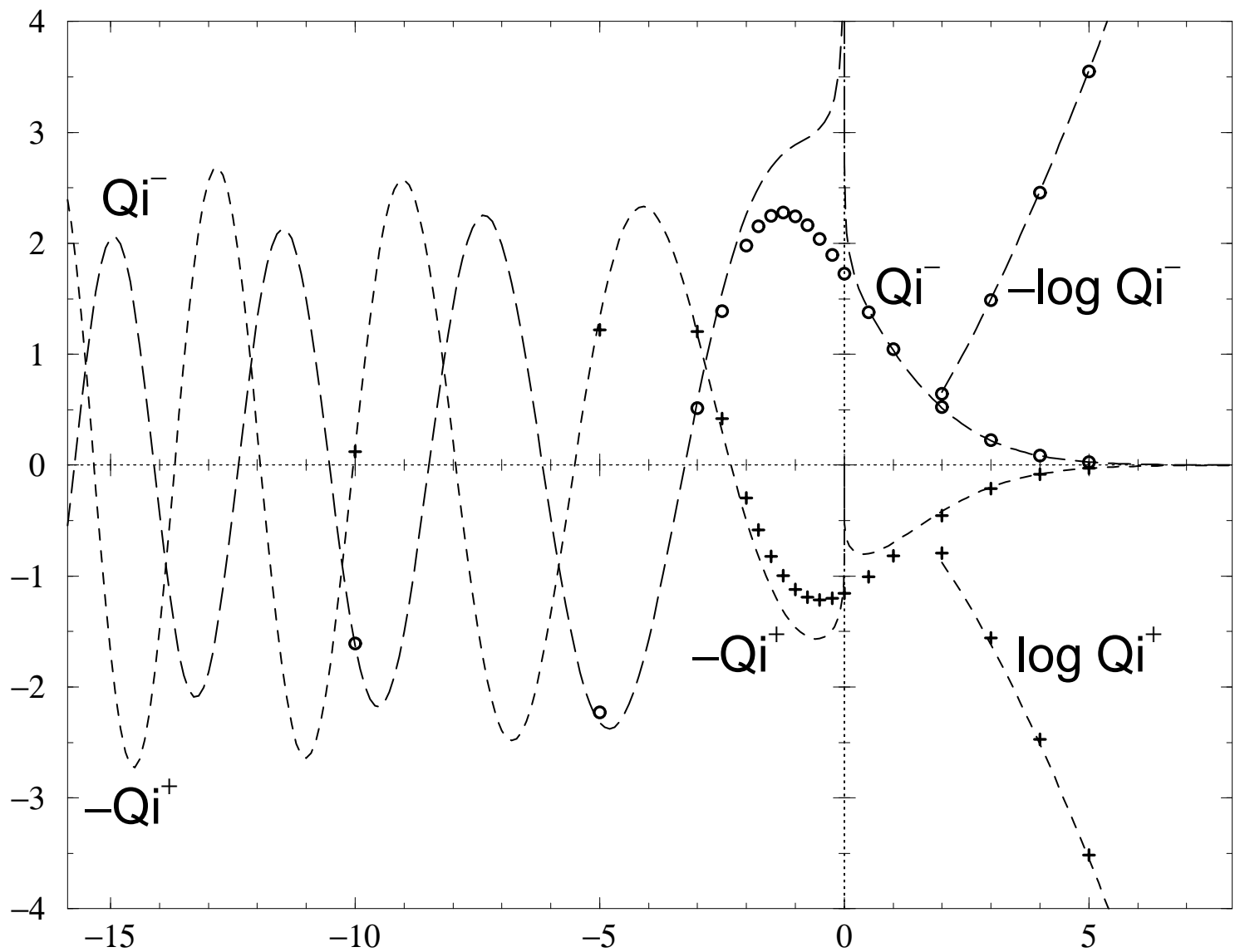
\end{document}